\documentclass[11pt,english,preprint]{aastex}

\usepackage{graphicx}  

\begin{document}
\title{The Properties of
Compressible MHD and Cosmic Ray Transport}

\author{A. Lazarian, Jungyeon Cho, \& Huirong Yan}

\affil{Dept. of Astronomy, Univ. of Wisconsin, Madison WI53706, USA}

\begin{abstract}
Turbulence is the most common state of astrophysical flows. In typical
astrophysical fluids, turbulence is accompanied by strong magnetic
fields, which has a large impact on the dynamics of the turbulent
cascade. Recently, there has been a significant breakthrough on the
theory of magnetohydrodynamic (MHD) turbulence. For the first time
we have a scaling model that is supported by both observations and
numerical simulations. We review recent progress in studies of both
incompressible and compressible turbulence. We compare Iroshnikov-Kraichnan
and Goldreich-Sridhar models, and discuss scalings of Alfv\'{e}n,
slow, and fast waves. We discuss the implications of this new insight
into MHD turbulence for cosmic ray transport. 
\end{abstract}

\section{Introduction}

Most astrophysical systems, e.g. accretion disks, stellar winds, the
interstellar medium (ISM) and intercluster medium are turbulent with
an embedded magnetic field that influences almost all of their properties.
This turbulence which spans from km to many kpc (see discussion in
Armstrong, Rickett, \& Spangler 1995; Scalo 1987; Lazarian, Pogosyan,
\& Esquivel 2002) holds the key to many astrophysical processes. For
instance, propagation of cosmic rays and their acceleration is strongly
affected by MHD turbulence. Recent research has shown that a substantial
part of the earlier results in the field require revision. Earlier
research used ad hoc models of MHD turbulence and this entailed erroneous
conclusions.

Before we start a discussion of MHD turbulence let us recall some
basic properties of the hydrodynamic turbulence. All turbulent systems
have one thing in common: they have a large {}``Reynolds number\char`\"{}
(\( Re\equiv LV/\nu  \); L= the characteristic scale or driving scale
of the system, V=the velocity difference over this scale, and \( \nu  \)=viscosity),
the ratio of the viscous drag time on the largest scales (\( L^{2}/\nu  \))
to the eddy turnover time of a parcel of gas (\( L/V \)). A similar
parameter, the {}``magnetic Reynolds number\char`\"{}, \( Rm \)
(\( \equiv LV/\eta  \); \( \eta  \)=magnetic diffusion), is the
ratio of the magnetic field decay time (\( L^{2}/\eta  \)) to the
eddy turnover time (\( L/V \)). The properties of the flows on all
scales depend on \( Re \) and \( Rm \). Flows with \( Re<100 \)
are laminar; chaotic structures develop gradually as \( Re \) increases,
and those with \( Re\sim10 ^{3} \) are appreciably less chaotic than
those with \( Re\sim10 ^{7} \). Observed features such as star forming
clouds and accretion disks are very chaotic with \( Re>10^{8} \)
and \( Rm>10^{16} \). 

Let us start by considering incompressible hydrodynamic turbulence,
which can be described by the Kolmogorov theory (Kolmogorov 1941).
Suppose that we excite fluid motions at a scale \( L \). We call
this scale the \textit{energy injection scale} or the \textit{largest
energy containing eddy scale}. For instance, an obstacle in a flow
excites motions on scales of the order of its size. Then the energy
injected at the scale \( L \) cascades to progressively smaller and
smaller scales at the eddy turnover rate, i.e. \( \tau _{l}^{-1}\approx v_{l}/l \),
with negligible energy losses along the cascade \footnote{%
This is easy to see as the motions at the scales of large eddies have
\( Re\gg 1 \).
}. Ultimately, the energy reaches the molecular dissipation scale \( l_{d} \),
i.e. the scale where the local \( Re\sim 1 \), and is dissipated
there. The scales between \( L \) and \( l_{d} \) are called the
\textit{inertial range} and it typically covers many decades. The
motions over the inertial range are \textit{self-similar} and this
provides tremendous advantages for theoretical description. 

The beauty of the Kolmogorov theory is that it does provide a simple
scaling for hydrodynamic motions. If the velocity at a scale \( l \)
from the inertial range is \( v_{l} \), the Kolmogorov theory states
that the kinetic energy (\( \rho v_{l}^{2}\sim v_{l}^{2} \) as the
density is constant) is transferred to next scale within one eddy
turnover time (\( l/v_{l} \)). Thus within the Kolmogorov theory
the energy transfer rate (\( v_{l}^{2}/(l/v_{l}) \)) is scale-independent,
\begin{equation}
\label{scale_{i}ndep}
\frac{v_{l}^{2}}{t_{cas}}\sim \frac{v_{l}^{2}}{(l/v_{l})}=\mbox {constant},
\end{equation}
 and we get the famous Kolmogorov scaling \begin{equation}
v_{l}\propto l^{1/3}.
\end{equation}

The one-dimensional\footnote{%
Dealing with observational data, e.g. in LPE02 (Lazarian et~al. 2002),
we deal with three dimensional energy spectrum \( P(k) \), which,
for isotropic turbulence, is given by \( E(k)=4\pi k^{2}P(k) \).
} energy spectrum \( E(k) \) is the amount of energy between the wavenumber
\( k \) and \( k+dk \) divided by \( dk \). When \( E(k) \) is
a power law, \( kE(k) \) is the energy \textit{near} the wavenumber
\( k\sim 1/l \). Since \( v_{l}^{2}\approx kE(k) \), Kolmogorov
scaling implies \begin{equation}
E(k)\propto k^{-5/3}.
\end{equation}

Kolmogorov scalings were the first major advance in the theory of
incompressible turbulence. They have led to numerous applications
in different branches of science (see Monin \& Yaglom 1975). However,
astrophysical fluids are magnetized and the a dynamically important
magnetic field should interfere with eddy motions. 

Paradoxically, astrophysical measurements are consistent with Kolmogorov
spectra (see LPE02). For instance, interstellar scintillation observations
indicate an electron density spectrum is very close to \( -5/3 \)
for \( 10^{8}cm \) - \( 10^{15}cm \) (see Armstrong et~al. 1995).
At larger scales LPE02 summarizes the evidence of -$5/3$  velocity
power spectrum over pc-scales in HI. Solar-wind observations provide
\textit{in-situ} measurements of the power spectrum of magnetic fluctuations
and Leamon et al.~ (1998) also obtained a slope of ~$\approx$ -$5/3$.
Is this a coincidence? What properties is the magnetized compressible
ISM expected to have? We will deal with these questions, and some
related issues, below. 

Here we discuss a focused approach which aims at obtaining a clear
understanding on the fundamental level, and considering physically
relevant complications later. The creative synthesis of both approaches
is the way, we think, that studies of astrophysical turbulence should
proceed\footnote{%
Potentially our approach leads to an understanding of the relationship
between motions at a given time at small scales (subgrid scales) and
the state of the flow at a previous time at some larger, resolved,
scale. This could lead to a parametrization of the subgrid scales
and to large eddy simulations of MHD.
}. Certainly an understanding of MHD turbulence in the most ideal terms
is a necessary precursor to understanding the complications posed
by more realistic physics and numerical effects. For review of general
properties of MHD, see a recent book by Biskamp (1993). 

In what follows, we first consider observational data that motivate
our study (\S2), then discuss theoretical approaches to incompressible
MHD turbulence (\S3). 
We move to the effects of compressibility in \S4 and discuss implications
of our new understanding of MHD turbulence for cosmic ray dynamics
in \S5. 
We present the summary in \S6.

\section{Observational Data}

Kolmogorov turbulence is the simplest possible model of turbulence.
Since it is incompressible and not magnetized, it is completely specified
by its velocity spectrum. If a passive scalar field, like {}``dye
particles{}'' or temperature inhomogeneities, is subjected to Kolmogorov
turbulence, the resulting spectrum of the passive scalar density is
also Kolmogorov (see Lesieur 1990; Warhaft 2000). In compressible
and magnetized turbulence this is no longer true, and a complete characterization
of the turbulence requires not only a study of the velocity statistics
but also the statistics of density and magnetic fluctuations. 

Direct studies of turbulence\footnote{%
Indirect studies include the line-velocity relationships (Larson 1981)
where the integrated velocity profiles are interpreted as the consequence
of turbulence. Such studies do not provide the statistics of turbulence
and their interpretation is very model dependent.
} have been done mostly for interstellar medium and for the Solar wind.
While for the Solar wind \textit{in-situ} measurements are possible,
studies of interstellar turbulence require inverse techniques to interpret
the observational data. 

Attempts to study interstellar turbulence with statistical tools date
as far back as the 1950s (von Horner 1951; Kamp\'{e} de F\'{e}riet
1955; Munch 1958; Wilson et~al. 1959) and various directions of research
achieved various degree of success (see reviews by Kaplan \& Pickelner
1970; Dickman 1985; Armstrong et~al. 1995; Lazarian 1999a, 1999b;
LPE02).

\subsection{Solar wind}

Solar wind (see review Goldstein \& Roberts 1995) studies allow pointwise
statistics to be measured directly using spacecrafts. These studies
are the closest counterpart of laboratory measurements. 

The solar wind flows nearly radially away from the Sun, at up to about
700 km/s. This is much faster than both spacecraft motions and the
Alfv\'{e}n speed. Therefore, the turbulence is {}``frozen{}'' and
the fluctuations at frequency \( f \) are directly related to fluctuations
at the scale \( k \) in the direction of the wind, as \( k=2\pi f/v \),
where \( v \) is the solar wind velocity (Horbury 1999). 

Usually two types of solar wind are distinguished, one being the fast
wind which originates in coronal holes, and the slower bursty wind.
Both of them show, however, \( f^{-5/3} \) scaling on small scales.
The turbulence is strongly anisotropic (see Klein et~al. 1993) with
the ratio of power in motions perpendicular to the magnetic field
to those parallel to the magnetic field being around 30. The intermittency
of the solar wind turbulence is very similar to the intermittency
observed in hydrodynamic flows (Horbury \& Balogh 1997).

\subsection{Electron density statistics}

Studies of turbulence statistics of ionized media (see Spangler \&
Gwinn 1990) have provided information on the statistics of plasma
density at scales \( 10^{8} \)-\( 10^{15} \)~cm. This was based
on a clear understanding of processes of scintillations and scattering
achieved by theorists\footnote{%
In fact, the theory of scintillations was developed first for the
atmospheric applications.
} (see Narayan \& Goodman 1989; Goodman \& Narayan 1985). A peculiar
feature of the measured spectrum (see Armstrong et~al. 1995) is the
absence of the slope change at the scale at which the viscosity by
neutrals becomes important. 

Scintillation measurements are the most reliable data in the {}``big
power law{}'' plot in Armstrong et al.~ (1995). However there are
intrinsic limitations to the scintillations technique due to the limited
number of sampling directions, its relevance only to ionized gas at
extremely small scales, and the impossibility of getting velocity
(the most important!) statistics directly. Therefore with the data
one faces the problem of distinguishing actual turbulence from static
density structures. Moreover, the scintillation data does not provide
the index of turbulence directly, but only shows that the data are
consistent with Kolmogorov turbulence. Whether the (3D) index can
be -4 instead of -11/3 is still a subject of intense debate (Higdon
1984; Narayan \& Goodman 1989). In physical terms the former corresponds
to the superposition of random shocks rather than eddies. 

Additional information on the electron density is contained in the
Faraday rotation measures of extragalactic radio sources (see Simonetti
\& Cordes 1988; Simonetti 1992). However, there is so far no reliable
way to disentangle contributions of the magnetic field and the density
to the signal. We feel that those measurements may give us the magnetic
field statistics when we know the statistics of electron density better.

\subsection{Velocity and density statistics from spectral lines}

Spectral line data cubes are unique sources of information on interstellar
turbulence. Doppler shifts due to supersonic motions contain information
on the turbulent velocity field which is otherwise difficult to obtain.
Moreover, the statistical samples are extremely rich and not limited
to discrete directions. In addition, line emission allows us to study
turbulence at large scales, comparable to the scales of star formation
and energy injection. 

However, the problem of separating velocity and density fluctuations
within HI data cubes is far from trivial (Lazarian 1995, 1999b; Lazarian
\& Pogosyan 2000; LPE02). The analytical description
of the emissivity statistics of channel maps (velocity slices) in
Lazarian \& Pogosyan (2000) (see also Lazarian 1999b; LPE02
for reviews) shows that the relative contribution of the density
and velocity fluctuations depends on the thickness of the velocity
slice. In particular, the power-law asymptote of the emissivity fluctuations
changes when the dispersion of the velocity at the scale under study
becomes of the order of the velocity slice thickness (the integrated
width of the channel map). These results are the foundation of the
Velocity-Channel Analysis (VCA) technique which provides velocity
and density statistics using spectral line data cubes. The VCA has
been successfully tested using data cubes obtained via compressible
magnetohydrodynamic simulations and has been applied to Galactic and
Small Magellanic Cloud atomic hydrogen (HI) data (Lazarian et~al.
2001; Lazarian \& Pogosyan 2000; Stanimirovic \& Lazarian 2001; Deshpande,
Dwarakanath, \& Goss 2000). Furthermore, the inclusion of absorption
effects (Lazarian \& Pogosyan 2002) has increased the power of this
technique. Finally, the VCA can be applied to different species (CO,
H\( _{\alpha } \) etc.) which should further increase its utility
in the future. 

Within the present discussion a number of results obtained with the
VCA are important. First of all, the Small Magellanic Cloud (SMC)
HI data exhibit a Kolmogorov-type spectrum for velocity and HI density
from the smallest resolvable scale of 40~pc to the scale of the SMC
itself, i.e. 4~kpc. Similar conclusions can be inferred from the
Galactic data (Green 1993) for scales of dozens of parsecs, although
the analysis has not been done systematically. Deshpande et al. (2000)
studied absorption of HI on small scales toward Cas A and Cygnus A.
Within the VCA their results can be interpreted as implying that on
scales less than 1~pc the HI velocity is suppressed by ambipolar
drag and the spectrum of density fluctuations is shallow \( P(k)\sim k^{-2.8} \).
Such a spectrum (Deshpande 2000) can account for the small scale structure
of HI observed in absorption.

\subsection{Magnetic field statistics}

Magnetic field statistics are the most poorly constrained aspect of
ISM turbulence. The polarization of starlight and of the Far-Infrared
Radiation (FIR) from aligned dust grains is affected by the ambient
magnetic fields. Assuming that dust grains are always aligned with
their longer axes perpendicular to magnetic field (see the review
Lazarian 2000), one gets the 2D distribution of the magnetic field
directions in the sky. Note that the alignment is a highly non-linear
process in terms of the magnetic field and therefore the magnetic
field strength is not available\footnote{%
The exception to this may be the alignment of small grains which can
be revealed by microwave and UV polarimetry (Lazarian 2000).
}. 

The statistics of starlight polarization (see Fosalba et~al. 2002)
is rather rich for the Galactic plane and it allows to establish the
spectrum\footnote{%
Earlier papers dealt with much poorer samples (see Kaplan \& Pickelner
1970) and they did not reveal power-law spectra.
} \( E(K)\sim K^{-1.5} \), where \( K \) is a two dimensional wave
vector describing the fluctuations over sky patch.\footnote{%
This spectrum is obtained by Fosalba et~al. (2002) in terms of the
expansion over the spherical harmonic basis \( Y_{lm} \). For sufficiently
small areas of the sky analyzed the multipole analysis results coincide
with the Fourier analysis.
} For uniformly sampled turbulence it follows from Lazarian \& Shutenkov
(1990) that \( E(K)\sim K^{\alpha } \) for \( K<K_{0} \) and \( K^{-1} \)
for \( K>K_{0} \), where \( K_{0}^{-1} \) is the critical angular
size of fluctuations which is proportional to the ratio of the injection
energy scale to the size of the turbulent system along the line of
sight. For Kolmogorov turbulence \( \alpha =-11/3 \). 

However, the real observations do not uniformly sample turbulence.
Many more close stars are present compared to the distant ones. Thus
the intermediate slops are expected. Indeed, Cho \& Lazarian (2002b)
showed through direct simulations that the slope obtained in Fosalba
et~al. (2002) is compatible with the underlying Kolmogorov turbulence.
At the moment FIR polarimetry does not provide maps that are really
suitable to study turbulence statistics. This should change soon when
polarimetry becomes possible using the airborne SOFIA observatory.
A better understanding of grain alignment (see Lazarian 2000) is required
to interpret the molecular cloud magnetic data where some of the dust
is known not to be aligned (see Lazarian, Goodman, \& Myers 1997 and
references therein). 

Another way to get magnetic field statistics is to use synchrotron
emission. Both polarization and intensity data can be used. The angular
correlation of polarization data (Baccigalupi et~al. 2001) shows
the power-law spectrum \( K^{-1.8} \) and we believe that the interpretation
of it is similar to that of starlight polarization. Indeed, Faraday
depolarization limits the depth of the sampled region. The intensity
fluctuations were studied in Lazarian \& Shutenkov (1990) with rather
poor initial data and the results were inconclusive. Cho \& Lazarian
(2002b) interpreted the fluctuations of synchrotron emissivity (Giardino
et~al. 2001, 2002) in terms of turbulence with Kolmogorov spectrum.

\section{Theoretical Approaches to MHD Turbulence}

Attempts to describe magnetic turbulence statistics were made by Iroshnikov
(1963) and Kraichnan (1965). Their model of turbulence (IK theory)
is isotropic in spite of the presence of the magnetic field. 

For simplicity, let us suppose that a uniform external magnetic field
(\( {\textbf {B}}_{0} \)) is present. In the incompressible limit,
any magnetic perturbation propagates \textit{along} the magnetic field
line. Since wave packets are moving along the magnetic field line,
there are two possible directions for propagation. If all the wave
packets are moving in one direction, then they are stable to nonlinear
order (Parker 1979). Therefore, in order to initiate turbulence, there
must be opposite-traveling wave packets and the energy cascade occurs
only when they collide. The IK theory starts from this observation,
one of the consequences of which is the modification of the energy
cascade timescale: \( t_{cas}\sim (L/V)(V_{A}/V) \), where
$V_A=B_0/\sqrt{4 \pi \rho}$ is Alfven velocity of the mean field.
Here, the IK
theory assumes that opposite-traveling isotropic wave packets of similar
size interact. From this and the scale-invariance of energy cascade
rate, they obtained \begin{equation}
\mbox {\textit {Iroshnikov-Kraichnan:}}E(k)\propto k^{-3/2}.
\end{equation}

However, the presence of the uniform magnetic component has non-trivial
dynamical effects on the turbulence fluctuations. One obvious effect
is that it is easy to mix field lines in directions perpendicular
to the local mean magnetic field and much more difficult to bend them.
The IK theory assumes isotropy of the energy cascade in Fourier space,
an assumption which has attracted severe criticism (Montgomery \&
Turner 1981; Shebalin, Matthaeus, \& Montgomery 1983; Montgomery \&
Matthaeus 1995; Sridhar \& Goldreich 1994; Matthaeus et~al. 1998).
Mathematically, anisotropy manifests itself in the resonant conditions
for 3-wave interactions: \begin{eqnarray}
{\textbf {k}}_{1}+{\textbf {k}}_{2} & = & {\textbf {k}}_{3},\label{k123} \\
\omega _{1}+\omega _{2} & = & \omega _{3},\label{w123} 
\end{eqnarray}
 where \( {\textbf {k}} \)'s are wavevectors and \( \omega  \)'s
are wave frequencies. The first condition is a statement of wave momentum
conservation and the second is a statement of energy conservation.
Alfv\'{e}n waves satisfy the dispersion relation: \( \omega =V_{A}|k_{\Vert }| \),
where \( k_{\Vert } \) is the component of wavevector parallel to
the background magnetic field. Since only opposite-traveling wave
packets interact, \( {\textbf {k}}_{1} \) and \( {\textbf {k}}_{2} \)
must have opposite signs. Then from equations (\ref{k123}) and (\ref{w123}),
either \( k_{\Vert ,1} \) or \( k_{\Vert ,2} \) must be equal to
0 and \( k_{\Vert ,3} \) must be equal to the nonzero initial parallel
wavenumber. That is, zero frequency modes are essential for energy
transfer (Shebalin et~al. 1983). Therefore, in the wavevector space,
3-wave interactions produce an energy cascade which is strictly perpendicular
to the mean magnetic field. However, in real turbulence, equation
(\ref{w123}) does not need to be satisfied exactly, but only to within
an an error of order \( \delta \omega \sim 1/t_{cas} \) (Goldreich
\& Sridhar 1995). This implies that the energy cascade is not strictly
perpendicular to \( {\textbf {B}}_{0} \), although clearly very anisotropic. 

We assume throughout this discussion that the rms turbulent velocity
at the energy injection scale is comparable to the Alfv\'{e}n speed
of the mean field and consider only scales below the energy injection
scale. This is called \textit{strong} turbulence regime. Note that,
as a consequence, the regime of \( B_{0}\gg \delta b \) is not considered
in this review. However, the regime of \( B_{0}\ll \delta b \) is
still relevant to the strong turbulence regime because scales below
the energy equipartition scale is expected to fall in the strong turbulence
regime (Cho \& Vishniac 2000a). 

An ingenious model very similar in its beauty and simplicity to the
Kolmogorov model has been proposed by Goldreich \& Sridhar
(1995; hereinafter GS95) for incompressible strong MHD turbulence.
They pointed out that motions perpendicular to the magnetic field
lines mix them on a hydrodynamic time scale, i.e. at a rate \( t_{cas}^{-1}\approx k_{\bot }v_{l} \),
where \( k_{\bot } \) is the wavevector component perpendicular to
the local mean magnetic field and \( l\sim k^{-1}(\approx k_{\perp }^{-1}) \).
These mixing motions couple to the wave-like motions parallel to magnetic
field giving a \textit{critical balance} condition \begin{equation}
\label{cr_{b}al}
k_{\Vert }V_{A}\sim k_{\bot }v_{k},
\end{equation}
 where \( k_{\Vert } \) is the component of the wavevector parallel
to the local magnetic field. When the typical \( k_{\Vert } \) on
a scale \( k_{\bot } \) falls below this limit, the magnetic field
tension is too weak to affect the dynamics and the turbulence evolves
hydrodynamically, in the direction of increasing isotropy in phase
space. This quickly raises the value of \( k_{\Vert } \). In the
opposite limit, when \( k_{\Vert } \) is large, the magnetic field
tension dominates, the error \( \delta \omega  \) in the matching
conditions is reduced, and the nonlinear cascade is largely in the
\( k_{\bot } \) direction, which restores the critical balance. 

If conservation of energy in the turbulent cascade applies locally
in phase space then the energy cascade rate (\( v_{l}^{2}/t_{cas} \))
is constant: \( (v_{l}^{2})/(l/v_{l})=\mbox {constant.} \)
Combining this with the critical balance condition we obtain an
anisotropy that increases with decreasing scale 
\begin{equation}
\label{twothirds}
k_{\Vert }\propto k_{\perp }^{2/3},
\end{equation}
 and a Kolmogorov-like spectrum for perpendicular motions \begin{equation}
v_{l}\propto l^{1/3},\mbox {or,}E(k)\propto k_{\perp }^{-5/3},
\end{equation}
 which is not surprising since the magnetic field does not influence
motions that do not bend it. At the same time, the scale-dependent
anisotropy reflects the fact that it is more difficult for the weaker,
smaller eddies to bend the magnetic field. 

GS95 shows the duality of motions in MHD turbulence. Those perpendicular
to the mean magnetic field are essentially eddies, while those parallel
to magnetic field are waves. The critical balance condition couples
these two types of motions. 

Numerical simulations (Cho \& Vishniac 2000b; Maron \& Goldreich 2001;
Cho, Lazarian, \& Vishniac 2002) show reasonable agreements with
the GS95 model.

\section{Compressible Turbulence}

\begin{table}[b!]
\caption{Notations for compressible turbulence}
\begin{center}
\setlength\tabcolsep{5pt}
\begin{tabular}{ll}
\hline\noalign{\smallskip}
 Notation   &  Meaning  \\
\noalign{\smallskip}
\hline
\noalign{\smallskip}
  a, $c_s$, $c_f$, $V_A$    &    sound, slow, fast, and Alfv\'en speed \\
  $\delta V$, $(\delta V)_s$, $(\delta V)_f$, $(\delta V)_A$
                            &  random (rms) velocity  \\
                            & Previously we used $V$ for the rms velocity \\
  $v_l$, $(v_l)_s$, $(v_l)_f$, $(v_l)_A$  &  velocity at scale $l$    \\
  ${\bf v}_{\bf k}$, $({\bf v}_{\bf k})_s$, $({\bf v}_{\bf k})_f$, 
  $({\bf v}_{\bf k})_A$       
                     &               velocity vector at wavevector ${\bf k}$ \\
  $\hat{\bf B}_0$ (=$\hat{\bf k}_{\|}$), $\hat{\bf k}_{\perp}$,
  $\hat{\bf k}$, $\hat{\bf \theta}$, ...      & unit vectors       \\
  ${\bf \xi}_s$, ${\bf \xi}_f$      & displacement vectors       \\
\hline
\end{tabular}
\end{center}
\label{cho_table1} 
\end{table}
For the rest of the review, we consider MHD turbulence
of a single conducting fluid.
While the GS95 model describes incompressible
MHD turbulence well,
no universally accepted theory exists 
for compressible MHD turbulence despite various
attempts (e.g., Higdon 1984).
Earlier numerical simulations of compressible MHD turbulence covered
a broad range of astrophysical problems, such as the decay of turbulence
(e.g.~{Mac Low} 1998; Stone, Ostriker, \& Gammie 1998) 
or
turbulent modeling of the ISM 
(see recent review Vazquez-Semadeni 2002; 
see also Passot, Pouquet, \& Woodward 1988; Vazquez-Semadeni, Passot, \&  Pouquet 1995; Passot, Vazquez-Semadeni, \& Pouquet 1995; Vazquez-Semadeni, Passot, \&  Pouquet 1996 for earlier pioneering 2D simulations
and Ostriker, Gammie, \& Stone 1999; Ostriker, Stone, \& Gammie 2001; Padoan {et~al.} 2001; Klessen 2001; Boldyrev 2002 for recent 3D simulations).
In what follows, we concentrate on the fundamental properties of
compressible MHD.

\subsection{Alfv\'en, slow, and fast modes}

\begin{figure}[t]
\begin{center}
\includegraphics{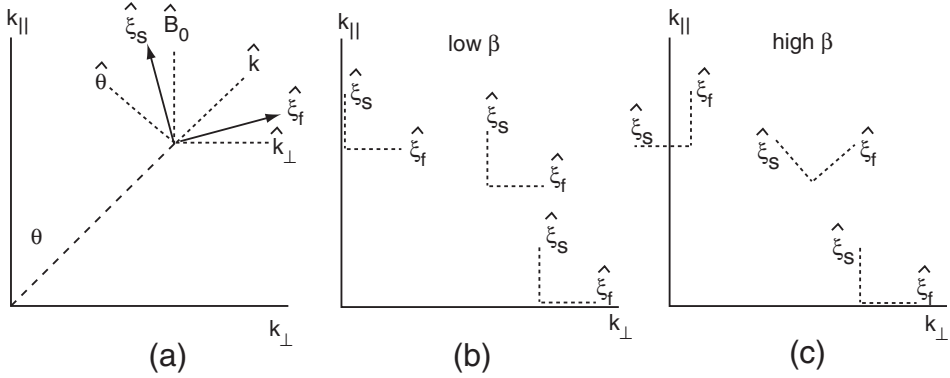} 
\end{center}
\caption{
         (a) Directions of fast and slow basis vectors.
             $\hat{\xi}_f$ and $\hat{\xi}_s$ represent
             the directions of displacement of fast and slow
             modes, respectively.
             In the fast basis ($\hat{\xi}_f$) is always between
             $\hat{\bf k}$ and $\hat{\bf k}_{\perp}$.
             In the 
             slow basis ($\hat{\xi}_s$) lies between
             $\hat{\theta}$ and $\hat{\bf B}_0$.
             Here, $\hat{\theta}$ is perpendicular to 
             $\hat{\bf k}$ and parallel to the wave front.
             All vectors lie in the same plane formed by
             ${\bf B}_0$ and ${\bf k}$.
             On the other hand, the displacement vector for
             Alfv\'en waves (not shown) is perpendicular to the plane.
         (b) Directions of basis vectors for a very small $\beta$ drawn in the
             same plane as in (a).
             The fast bases are almost parallel to $\hat{\bf k}_{\perp}$.
         (c) Directions of basis vectors for a very high $\beta$.
             The fast basis vectors are almost parallel to ${\bf k}$.
             The slow waves become pseudo-Alfv\'en waves.
       }
\label{fig_modes}
\end{figure}  
\begin{figure}[t]
\begin{center}
\includegraphics{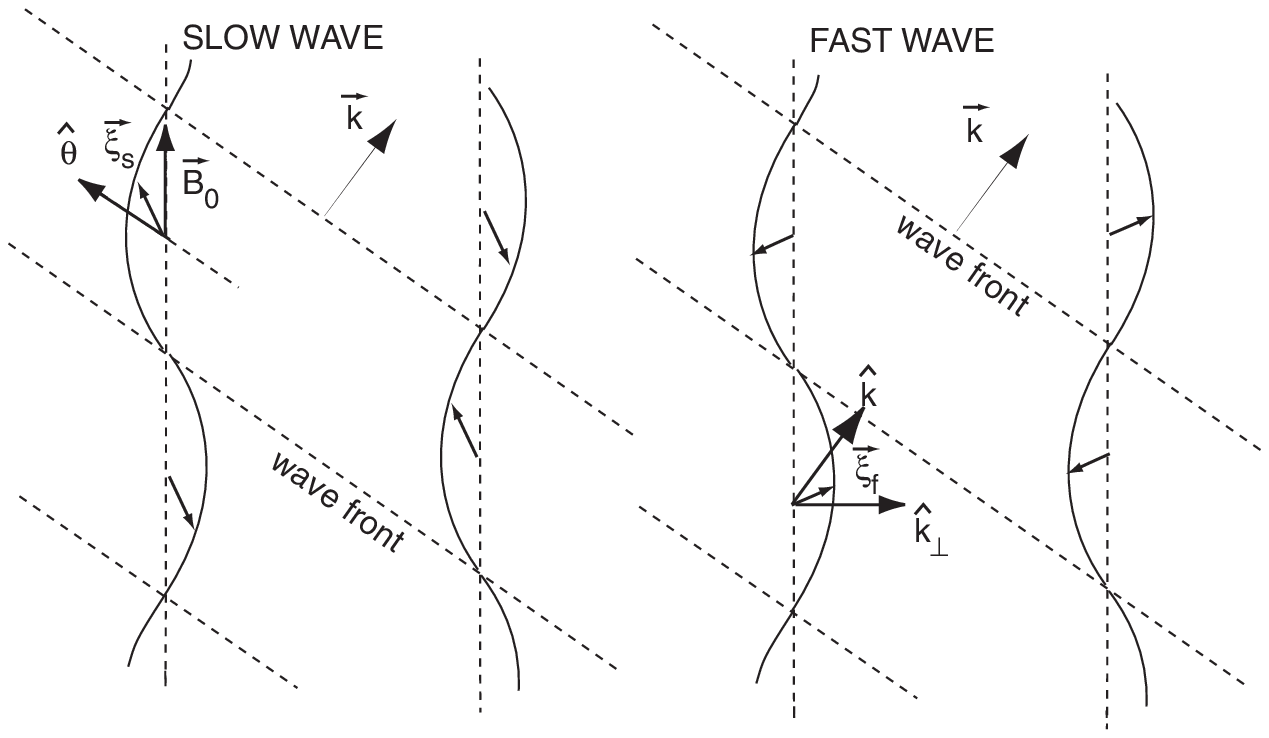}
\end{center}
\caption{
         Waves in real space.
         We show the directions of displacement vectors for
         a slow wave ({\it left}) and a fast wave ({\it right}).
         Note that $\hat{\xi}_s$ lies between
        $\hat{\theta}$ and $\hat{\bf B}_0 ~(=\hat{\bf k}_{\|})$ and 
                   $\hat{\xi}_f$ between
        $\hat{\bf k}$ and $\hat{\bf k}_{\perp}$.
         Again, $\hat{\theta}$ is perpendicular to 
             $\hat{\bf k}$ and parallel to the wave front.
         Note also that, for the fast wave, for example,
         density (inferred by the
         directions of the displacement vectors)
         becomes higher where field lines are closer, resulting
         in a strong restoring force, which is why fast waves
         are faster than slow waves.
       }
\label{fig_modes-real}
\end{figure}  
  
Let us start by reviewing different MHD waves.
In particular, we describe the Fourier space representation
of these waves.  The
real space representation can be found in   
papers on modern shock-capturing MHD codes  
(e.g.~Brio \& Wu 1988; Ryu \& Jones 1995).
{}For the sake of simplicity, we consider an isothermal plasma.
{}Figure \ref{fig_modes} and Figure \ref{fig_modes-real} give
schematics of slow and fast waves.
{}For slow and fast waves,
${\bf B}_0$, ${\bf v}_{\bf k}$ ($\propto \xi$), and ${\bf k}$ are
in the same plane.
On the other hand, for Alfv\'en waves, the velocity of the
{}fluid element $({\bf v}_{\bf k})_A$ is orthogonal to the 
${\bf B}_0 - {\bf k}$ plane.

As before, the Alfv\'en speed is
$
  V_A=B_0/\sqrt{4\pi \rho_0},
$
where $\rho_0$ is the average density.
{}Fast and slow speeds are
\begin{equation}
 c_{f,s} = \left[   \frac{1}{2} \left\{
         a^2+V_A^2 \pm \sqrt{ (a^2+V_A^2)^2 - 4 a^2 V_A^2 \cos^2{\theta} }
                                \right\} 
           \right]^{1/2},
\end{equation} 
where $\theta$ is the angle between ${\bf B}_0$ and ${\bf k}$.
See Table \ref{cho_table1} for the definition of other variables.
When $\beta$ 
($\beta=P_{g}/P_{B}$=$2a^2/V_A^2$; 
$P_g$= gas pressure, $P_B$= magnetic pressure;
hereinafter $\beta=$ average $\beta \equiv \bar{P}_{g}/\bar{P}_{B}$) 
goes to zero, we have
\begin{eqnarray} 
 c_f &\approx& V_A, \nonumber \\     
 c_s &\approx& a\cos{\theta}.   
\end{eqnarray}

Figure \ref{fig_modes} shows directions of displacement 
(or, directions of velocity) vectors
for these three modes.
We will call them the basis vectors for these modes.
The Alfv\'en basis is perpendicular to both $\hat{\bf k}$ 
and $\hat{\bf B}_0$,
and coincides with the azimuthal vector $\hat{\bf \phi}$ in 
a spherical-polar coordinate system.
Here hatted vectors are unit vectors.
The fast basis $\hat{\bf \xi}_f$ 
lies {\it between} $\hat{\bf k}$ and $\hat{\bf k}_{\perp}$:
\begin{equation}
   \hat{\bf \xi}_f \propto 
     \frac{ 1-\sqrt{D}+{\beta}/2  }{ 1+\sqrt{D}-{\beta}/2  } 
    \left[ \frac{ k_{\perp} }{ k_{\|} } \right]^2
     k_{\|} \hat{\bf k}_{\|}  +
          k_{\perp} \hat{\bf k}_{\perp},
\end{equation}    
where
$
  D=(1+{\beta}/2)^2-2{\beta} \cos^2{\theta}$, and
$  {\beta}$ is the averaged $\beta$ (=$\bar{P}_g/\bar{P}_B$).
The slow basis $\hat{\bf \xi}_s$ lies {\it between} $\hat{\bf \theta}$ 
and $\hat{\bf B}_0$ (=$\hat{\bf k}_{\|}$):
\begin{equation}
   \hat{\bf \xi}_s \propto 
        k_{\|} \hat{\bf k}_{\|}+
     \frac{ 1-\sqrt{D}-{\beta}/2  }{ 1+\sqrt{D}+{\beta}/2  } 
    \left[ \frac{ k_{\|} }{ k_{\perp} }  \right]^2
     k_{\perp} \hat{\bf k}_{\perp}. 
\end{equation}
The two vectors $\hat{\bf \xi}_f$ and $\hat{\bf \xi}_s$ are
mutually orthogonal.
Proper normalizations are required for both bases 
to make them unit-length.

When $\beta$ goes to zero (i.e. the magnetically dominated regime),
$\hat{\bf \xi}_f$ becomes parallel to $\hat{\bf k}_{\perp}$
and $\hat{\bf \xi}_s$ becomes parallel to $\hat{\bf B}_0$ 
(Fig.~\ref{fig_modes}b).
The sine of the angle between $\hat{\bf B}_0$ and $\hat{\bf \xi}_s$ 
is $(\beta/2) \sin\theta \cos{\theta}$.
When $\beta$ goes to infinity (i.e. gas pressure dominated regime)\footnote{
In this section, we assume that external mean field is strong 
(i.e.~$V_A > (\delta V)$) but finite, so that 
$\beta \rightarrow \infty$ 
means the gas pressure $\bar{P}_g \rightarrow \infty$. 
},
$\hat{\bf \xi}_f$ becomes parallel to $\hat{\bf k}$
and $\hat{\bf \xi}_s$ becomes parallel to $\hat{\bf \theta}$ 
(Fig.~\ref{fig_modes}c).
This is the incompressible limit.
In this limit, the slow mode is sometimes called the pseudo-Alfv\'en mode 
(Goldreich \& Sridhar 1995).

\subsection{Theoretical considerations}
Here we address the issue of mode coupling in a low $\beta$
plasma. It is reasonable to suppose that in the limit where $\beta \gg 1$
turbulence for Mach numbers ($M_s=\delta V/a$) less than unity
should be largely similar to the exactly incompressible regime.
Thus, Lithwick \& Goldreich  (2001) conjectured that the GS95 relations
are applicable to slow and Alfv\'en modes with the fast modes decoupled.
They also mentioned that this relation can carry on
for low $\beta$ plasmas.
{}For $\beta \gg1$ and $M_s>1$, we are in the regime of hydrodynamic 
compressible turbulence for which no theory exists, as far as we know.

In the diffuse interstellar medium $\beta$  is typically less than
unity. 
In addition, it is $\sim 0.1$ or less for molecular clouds.
There are a few simple arguments suggesting that MHD theory can be formulated 
in the regime where the Alfv\'en Mach number ($\equiv \delta V/V_A$) is
less than unity, although this is not a universally accepted
assumption.
Alfv\'en modes describe incompressible motions.
Arguments in GS95 are suggestive that the coupling of Alfv\'{e}n
to fast and slow modes
will be weak.
Consequently, we expect that in this regime the Alfv\'en
cascade should follow the GS95 scaling.
Moreover the slow waves are likely to evolve passively (Lithwick \& Goldreich 2001).
{}For $a \ll V_A$ their nonlinear evolution should be governed 
by Alfv\'en modes 
so that we expect the GS95 scaling for them as well.  The phase velocity of 
Alfv\'en waves and slow waves depend on a factor of $\cos{\theta}$ and
this enables modulation of the slow waves by the Alfv\'en ones.
However, fast waves do not have this factor and therefore cannot be
modulated by the changes of the magnetic field direction associated
with Alfv\'en waves. The coupling between the modes is through
the modulation of the local Alfv\'en velocity and therefore
is weak.

{}For Alfv\'en Mach number ($M_A$) larger than unity a
shock-type regime is expected.
However, generation of magnetic field by turbulence (Cho \& Vishniac 2000a) is expected
for such a regime.
It will make the steady state turbulence approach $M_A \sim 1$.\footnote{
We suspect that simulations 
that show super-Alfv\'{e}nic turbulence is widely spread in the ISM 
might not evolve for a long
enough time to reach the steady state.}
 Therefore in 
Cho \& Lazarian  (2002a) we consider turbulence in the limit
$M_s>1$, $M_A<1$, and $\beta <1$.
{}For these simulations, we mostly used $M_s\sim 2.2$, $M_A \sim 0.7$, and
$\beta \sim 0.2$. The Alfv\'en speed of the mean external field
is similar to the rms velocity ($V_A=1,\delta V\sim 0.7, a=\sqrt{0.1}$),
and we used an isothermal equation of state.

Although the scaling relations presented below are applicable to
sub-Alfv\'{e}nic turbulence, we cautiously speculate that
small scales of super-Alfv\'{e}nic turbulence
might follow similar scalings.
Boldyrev, Nordlund, \& Padoan  (2001) obtained energy spectra
close to $E(k)\sim k^{-1.74}$ in solenoidally driven super-Alfv\'{e}nic
supersonic turbulence simulations.
The spectra are close to the Kolmogorov spectrum ($\sim k^{-5/3}$), rather than
shock-dominated spectrum ($\sim k^{-2}$).
This result might imply that small scales of super-Alfv\'{e}nic MHD turbulence
can be described by our sub-Alfv\'{e}nic model presented below, which predicts
Kolmogorov-type spectra for Alfv\'{e}n and slow modes.

\subsection{Coupling of MHD modes and Scaling of Alfv\'{e}n modes} 
\label{sec_coupling}

Alfv\'en modes are not susceptible to collisionless damping (see
Spangler 1991; Minter \& Spangler 1997 and references therein) that damps slow and fast modes.
Therefore, we mainly consider the transfer of energy from  
Alfv\'en waves to compressible MHD waves (i.e. to the 
slow and fast modes).

\begin{figure}[t]
\includegraphics[width=.5\columnwidth]{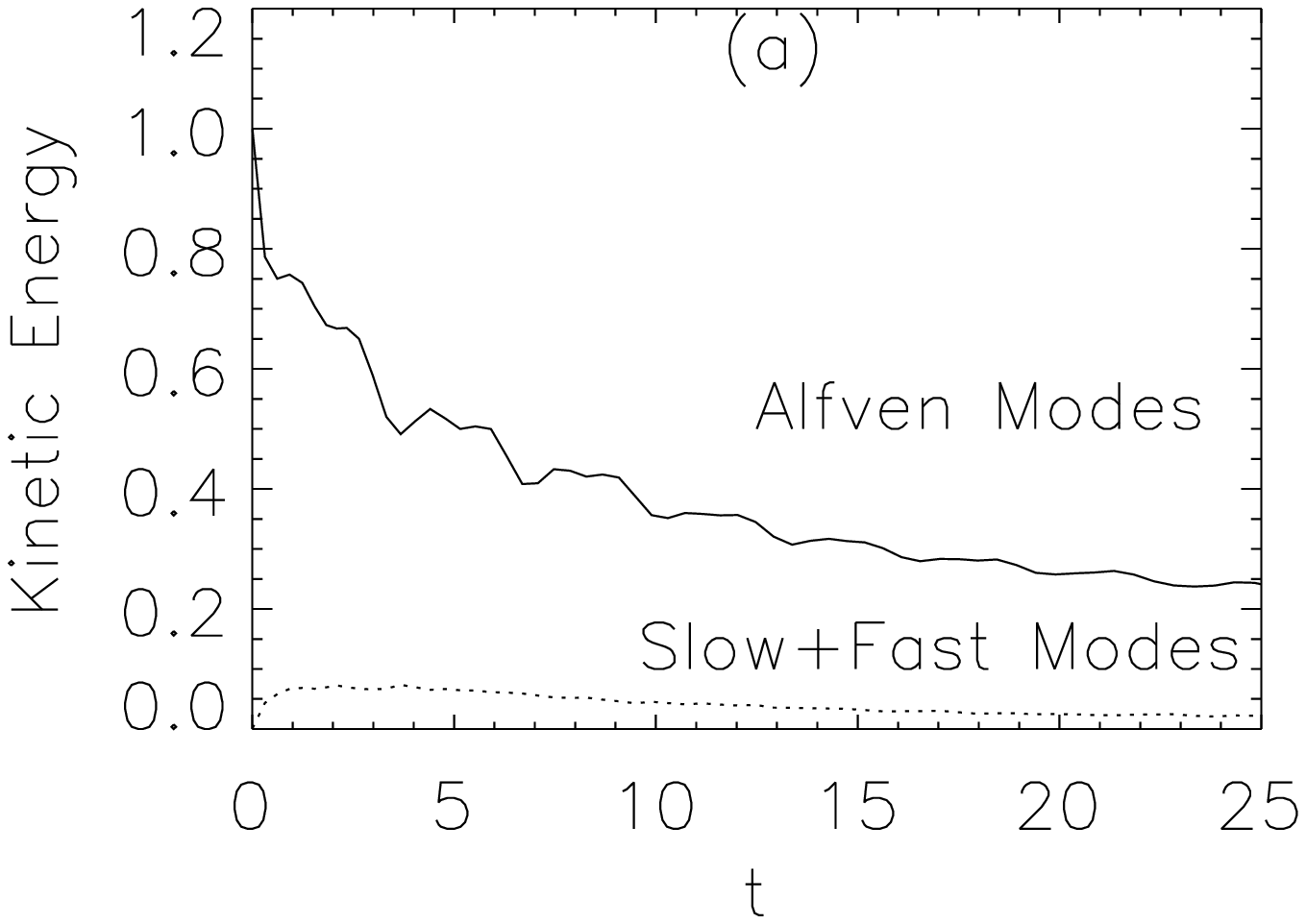}
\hfill
\includegraphics[width=.45\columnwidth]{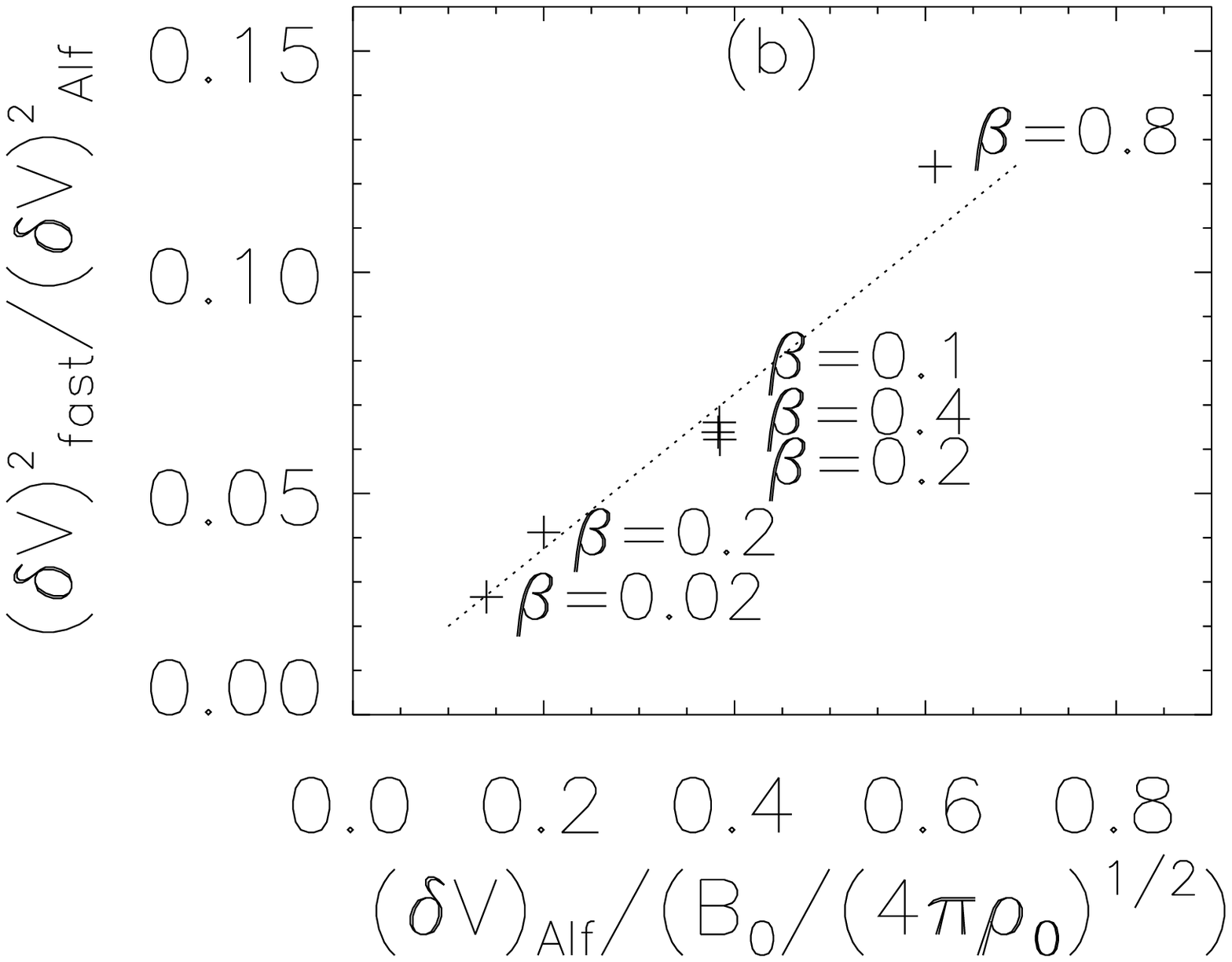}
\caption{
         ({\it Left}) Decay of Alfv\'{e}nic turbulence.
         The generation of fast and slow waves is not efficient.
         $\beta \sim 0.2$, $M_s \sim 3$.
         ({\it Right}) The ratio of $(\delta V)_f^2$ to
          $(\delta V)_A^2$. The stronger the external field ($B_0$) is,
          the more suppressed the coupling is.
          The ratio is not sensitive to $\beta$.
          {}From Cho \& Lazarian (2002a)
       }
\label{fig_coupling}
\end{figure}  

In Cho \& Lazarian  (2002a), we carry out simulations 
to check the strength of the mode-mode coupling.
We first obtain a data cube from a driven compressible numerical simulation with
$B_0/\sqrt{4 \pi \rho_0}=1$.
Then, after turning off the driving force, 
we let the turbulence decay.
We go through
the following procedures before we let the turbulence decay.
We first remove slow and fast modes in Fourier space and
retain only Alfv\'{e}n modes.
We also change the value of ${\bf B}_0$ preserving its original
direction.
We use the same constant initial density $\rho_0$ for all simulations.
We assign a new constant initial gas pressure $P_g$
 \footnote{
      The changes of both $B_0$ and $P_g$ preserve the 
      Alfv\'{e}n character of perturbations.
      In Fourier space,
      the mean magnetic field (${\bf B}_0$)
      is the amplitude of ${\bf k}={\bf 0}$ component.
      Alfv\'{e}n components in Fourier space are
      for ${\bf k} \neq {\bf 0}$ and their directions are 
      parallel/anti-parallel to
      $\hat{\bf \xi}_A$ (= $\hat{\bf B}_{0} \times \hat{\bf k}_{\perp}$).
      The direction of $\hat{\bf \xi}_A$ does not depend on
      the magnitude of $B_0$ or $P_g$.
}.
After doing all these procedures, we let the turbulence decay.
We repeat the above procedures for different values of $B_0$ and $P_g$.
{}Fig.~\ref{fig_coupling}a shows the evolution of the kinetic 
energy of a simulation.
The solid line represents the kinetic energy of Alfv\'en  modes.
It is clear that Alfv\'en waves are poorly coupled to the compressible
modes, and do not generate them efficiently \footnote{
As correctly pointed out by Zweibel (this volume) there is always 
residual coupling between Alfv\'{e}n and compressible modes due to
steepening of Alfv\'{e}n modes.
However, this steepening happens on time-scales much longer than the
cascading time-scale.}
Therefore, we expect that Alfv\'en modes will follow the same scaling
relation as in the incompressible case.
{}Fig.~\ref{fig_coupling}b shows that the coupling gets weaker as 
$B_0$ increases: 
\begin{equation}
  \frac{ (\delta V)_{f}^2 }{ (\delta V)_{A}^2 }
 \propto \frac{ (\delta V)_{A} }{ B_0 }. \label{vfscale}
\end{equation}
The ratio of $(\delta V)_{s}^2$ to $(\delta V)_{A}^2$ is proportional to
$(\delta V)_{A}^2/ B_0^2$.

\begin{figure}[!t]
\includegraphics[width=.5\columnwidth]{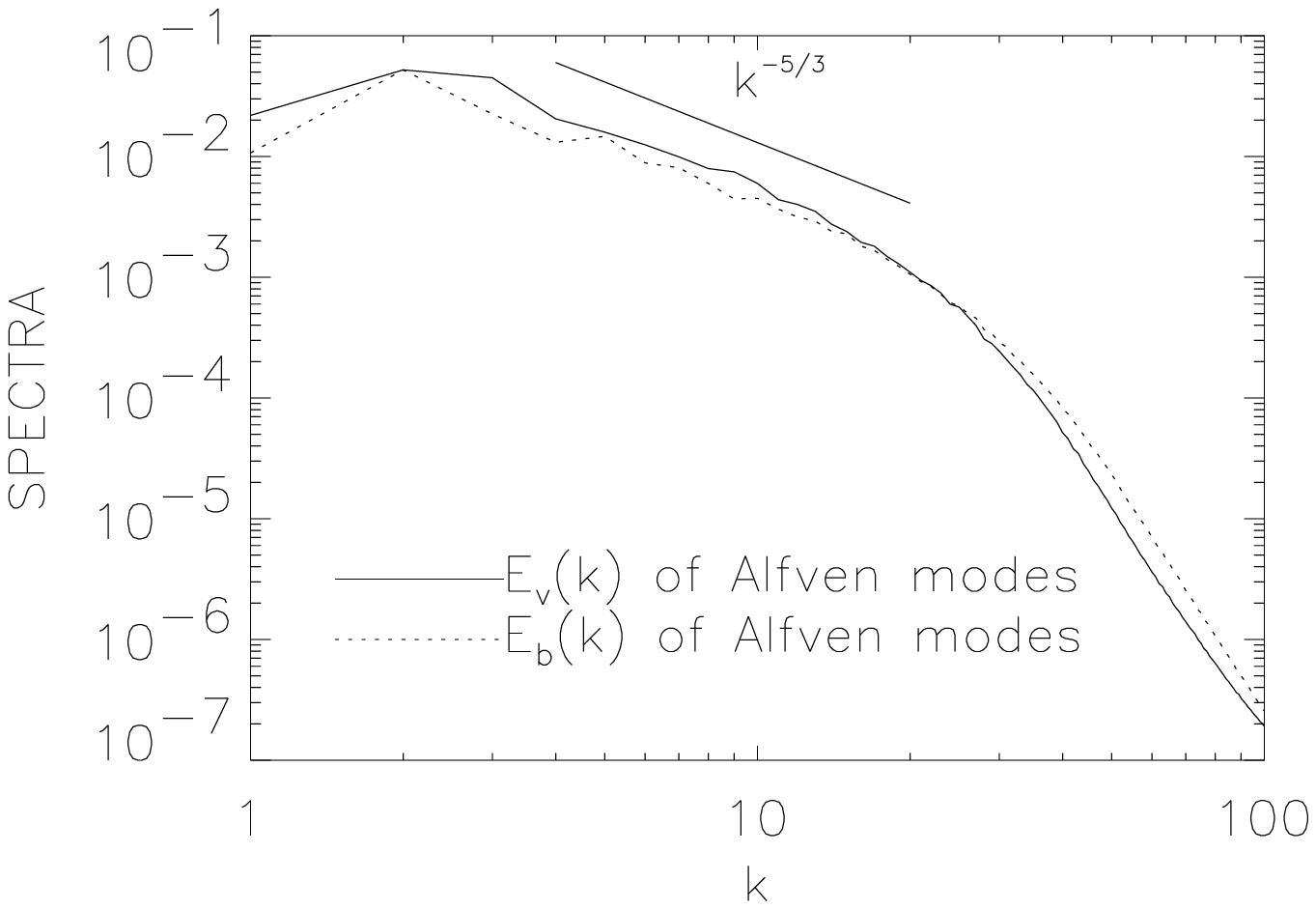}
\hfill
\includegraphics[width=.45\columnwidth]{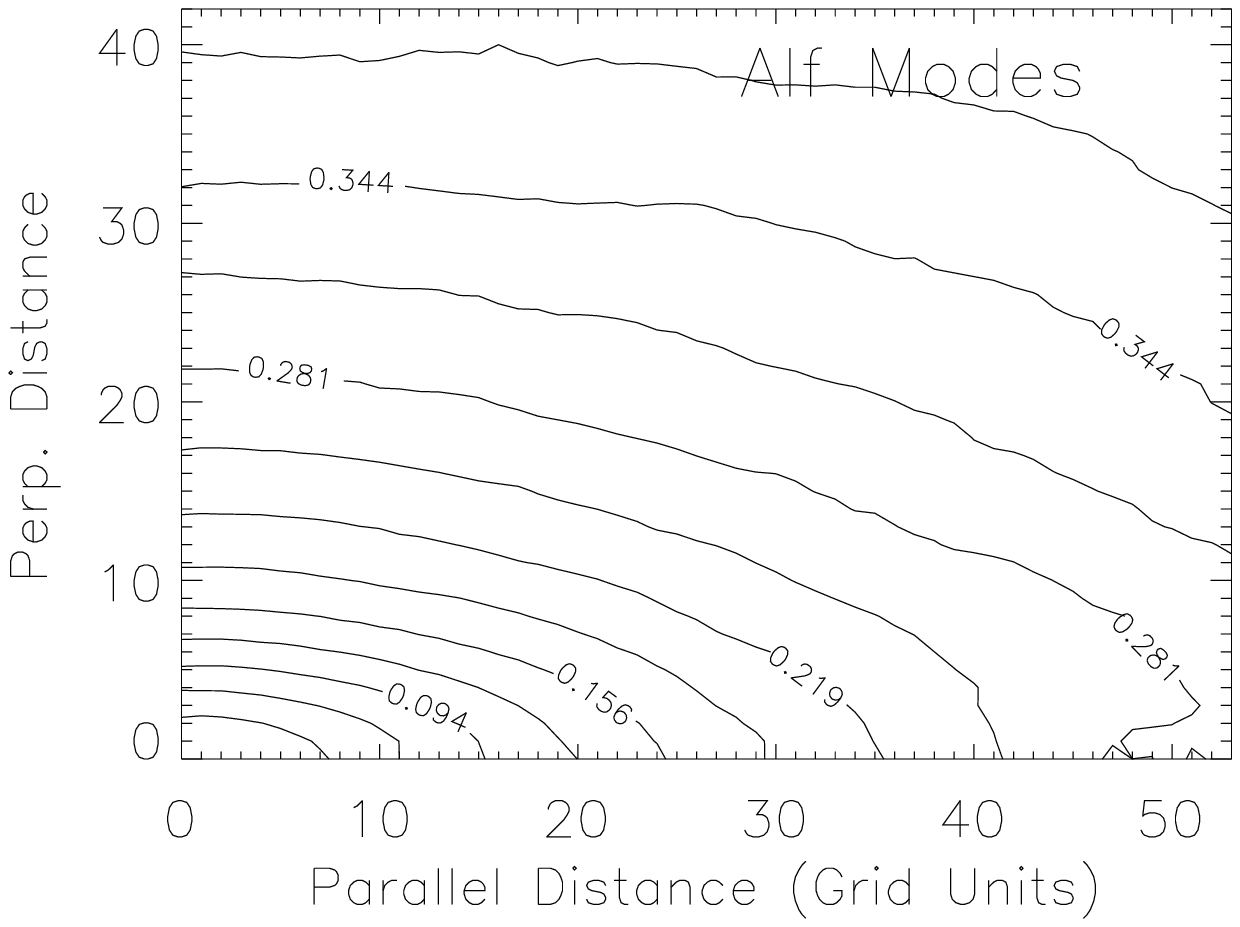}
\caption{
         (a) Alfv\'en spectra follow a Kolmogorov-like power law.
         (b) The second-order structure function 
             ($SF_2=<{\bf v}({\bf x}+{\bf r}) - {\bf v}({\bf x})>$) 
             for Alf\'ven velocity
             shows anisotropy similar to the GS95.
             Conturs represent eddy shapes.
             {}From Cho \& Lazarian (2002a).
       }
\label{fig_alf} 
\end{figure}  

This marginal coupling is in good agreement with a claim in GS95, as well as
earlier numerical studies where the
velocity was decomposed into a compressible component ${\bf v}_C$
and a solenoidal component ${\bf v}_S$. The compressible component 
is curl-free and parallel to the wave vector ${\bf k}$ in Fourier space.
The solenoidal component is divergence-free and perpendicular to ${\bf k}$.
The ratio $\chi = (\delta V)_C/ (\delta V)_S$ is an important parameter 
that determines the
strength of any shock
(Passot {et~al.} 1988; Pouquet 1999).
Porter, Woodward, \& Pouquet  (1998) performed a 
hydrodynamic simulation of
decaying turbulence with an initial sonic Mach number of unity 
and found that $\chi^2$ evolves toward $\sim 0.11$.
Matthaeus et al.~ (1996) 
carried out simulations of decaying weakly compressible
MHD turbulence (Zank \& Matthaeus 1993) and found that $\chi^2 \sim O(M_s^2)$, where
$M_s$ is the sonic Mach number.
In Boldyrev {et~al.} (2001) a weak generation of
compressible components in solenoidally driven super-Alfv\'{e}nic supersonic
turbulence simulations was obtained.

{}Fig.~\ref{fig_alf} shows that the spectrum and 
the anisotropy of Alfv\'en waves in this limit are
compatible with the GS95 model:
\begin{equation}
  \mbox{\it Spectrum of Alfv\'{e}n Modes:~~~~~} E(k)\propto k_{\perp}^{-5/3},
\end{equation}
and scale-dependent anisotropy 
$k_{\|}\propto k_{\perp}^{2/3}$ that is compatible with the GS95 theory.

\subsection{Scaling of the slow modes} \label{sec_slow}
Slow waves are somewhat 
similar to pseudo-Alfv\'en waves (in the incompressible limit).
{}First, the directions of displacement (i.e.~$\xi_s$) of both waves
are similar when anisotropy is present.
The vector ${\bf \xi}_s$ is always between $\hat{\bf \theta}$ and
$\hat{\bf k}_{\|}$.
In Figure \ref{fig_modes}, we can see that the angle between
$\hat{\bf \theta}$ and
$\hat{\bf k}_{\|}$  
gets smaller when $k_{\|} \ll k_{\perp}$.
Therefore, when there is anisotropy (i.e.~$k_{\|} \ll k_{\perp}$), 
$\hat{\bf \xi}_s$ of a low $\beta$ plasma becomes similar to 
that of a high $\beta$ plasma.
Second, the angular dependence in
the dispersion relation $c_s \approx a \cos{\theta}$ is identical to
that of pseudo-Alfv\'en waves (the only difference is that,
in slow waves, the sound speed $a$ is present instead of the Alfv\'en
speed $V_A$).

\begin{figure}[!t]
\includegraphics[width=.5\columnwidth]{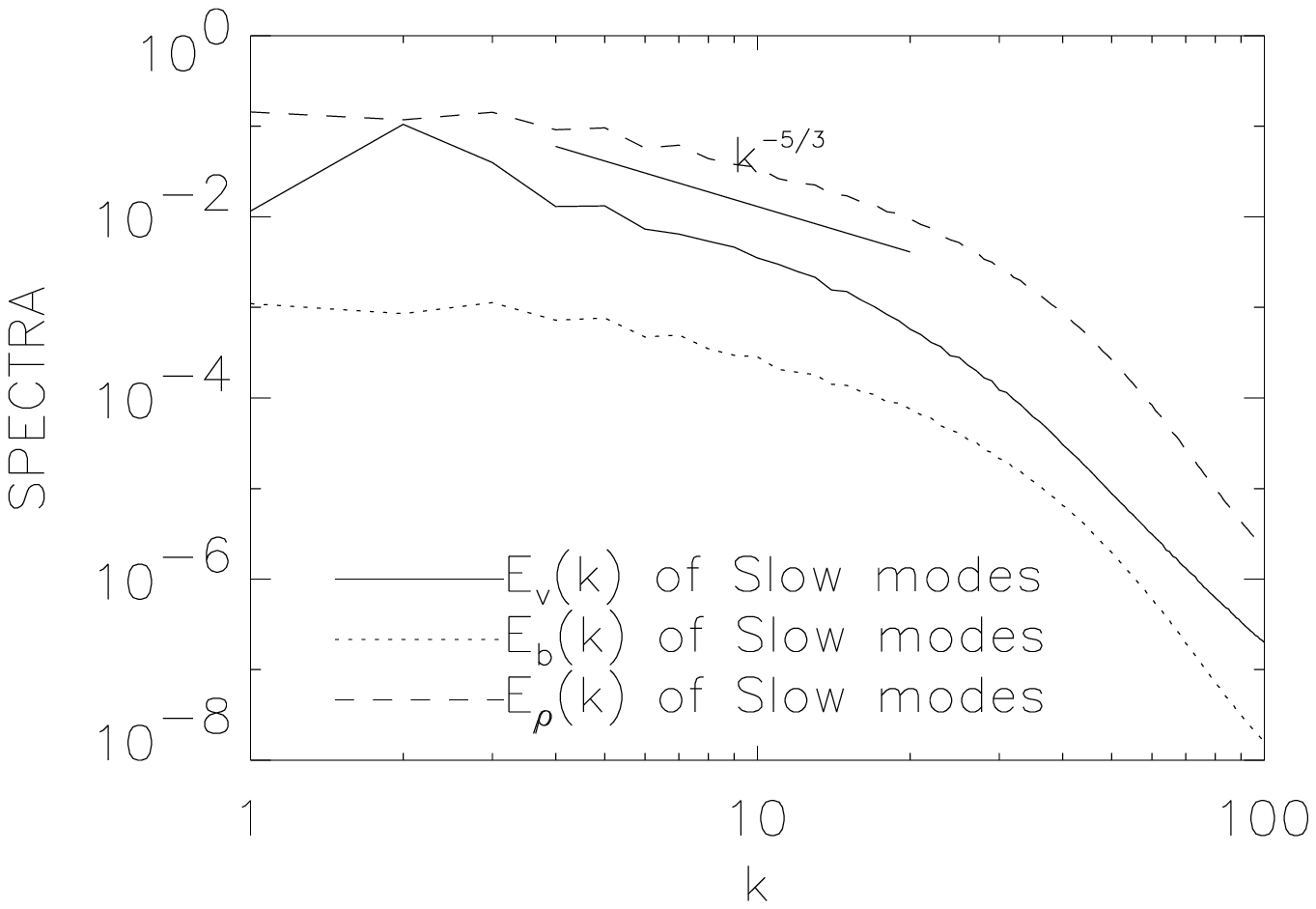}
\hfill
\includegraphics[width=.45\columnwidth]{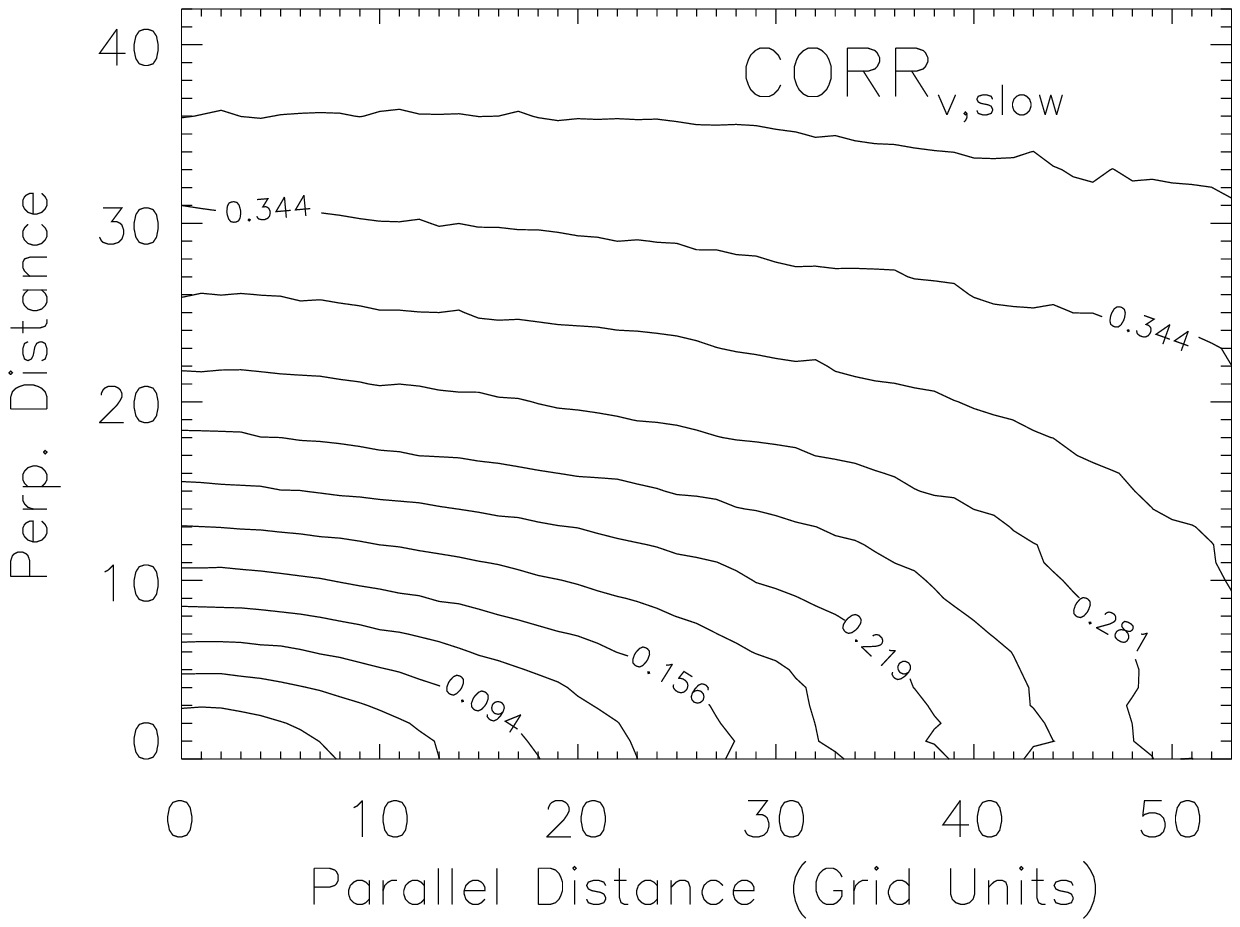}
\caption{   
         (a) Slow spectra also follow a Kolmogorov-like power law.
         (b) Slow modes show anisotropy similar to the GS95 theory.
             {}From Cho \& Lazarian (2002a).
       }
\label{fig_slow}
\end{figure}  

Goldreich \& Sridhar  (1997) argued that
the pseudo-Alfv\'en waves are slaved to the shear-Alfv\'en 
(i.e.~ordinary Alfv\'en)
waves in the presence of a strong ${\bf B}_0$, meaning
that the energy cascade of pseudo-Alfv\'en modes is primarily mediated
by the shear-Alfve\'en waves.
This is because pseudo-Alfv\'en waves do not provide efficient shearing
motions.
Similar arguments are applicable to slow waves
in a low $\beta$ plasma (Cho \& Lazarian 2002a) 
(see also Lithwick \& Goldreich 2001 for high-$\beta$ plasmas).
As a result, we conjecture that slow modes follow a scaling similar to
the GS95 model (Cho \& Lazarian 2002a):
\begin{equation}
 \mbox{\it Spectrum of Slow Modes:~~~~~}  E^{s}(k) \propto k_{\perp}^{-5/3}.
\end{equation}

{}Fig.~\ref{fig_slow}a shows the spectra of slow modes.
For velocity, the slope is close to $-5/3$.
Fig.~\ref{fig_slow}b shows the contours of equal 
second-order structure function ($SF_2$)
of slow velocity, which are compatible with $k_{\|}\propto k_{\perp}^{2/3}$
scaling.

In low $\beta$ plasmas, density fluctuations are dominated by
slow waves (Cho \& Lazarian 2002a).
{}From the continuity equation $\dot{\rho} = \rho \nabla \cdot {\bf v}$
\begin{equation}
  \omega \rho_k = \rho_0 {\bf k} \cdot  {\bf v}_k,
\end{equation}
we have,
{}for slow modes, 
$
(\rho_k)_s \sim \rho_0 (v_k)_s/a. \label{roks}
$
Hence, this simple argument implies
\begin{equation}
 \left(\frac{ \delta \rho }{ \rho }\right)_s = \frac{ (\delta V)_s }{ a }\sim  
             M_s,   \label{eq39}
\end{equation}
where we assume that $(\delta V)_s \sim (\delta V)_A$ and
$M_s$ is the Mach number.
On the other hand, only a small amount of magnetic field is produced by the
slow waves. Similarly, using the induction equation 
($\omega {\bf b}_k = {\bf k} \times ({\bf B}_0 \times {\bf v}_k)$), we have
\begin{equation}
 \frac{(\delta B)_s}{ (\delta V)_s } \sim \frac{a}{ B_0 } 
                            =O(\sqrt{\beta}),
\end{equation}
which means that equipartition between kinetic and magnetic energy
is not guaranteed in low $\beta$ plasmas.
In fact, in Fig.~\ref{fig_slow}a, the power spectrum 
for density fluctuations has a much larger amplitude than 
the magnetic field power spectrum.
Since density fluctuations are caused mostly by the slow waves and
magnetic fluctuation is caused mostly by Alfv\'{e}n and fast modes,
we {\it do not} expect a strong correlation between density and
magnetic field, which agrees with the ISM simulations
(Padoan \& Nordlund 1999; Ostriker {et~al.} 2001; Vazquez-Semadeni 2002).

\subsection{Scaling of the fast modes} 
\begin{figure}[!t]
\includegraphics[width=.5\columnwidth]{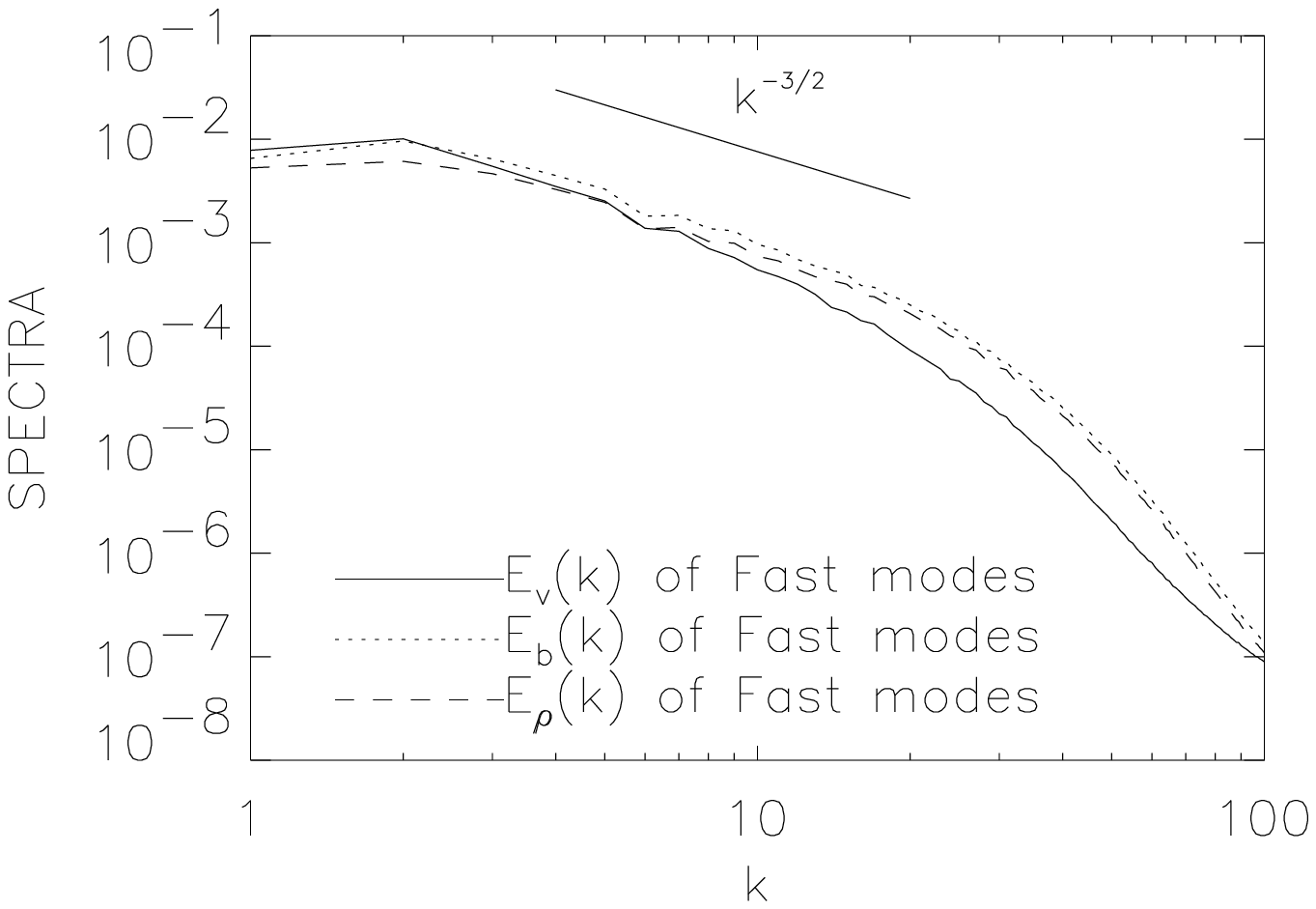}
\hfill
\includegraphics[width=.45\columnwidth]{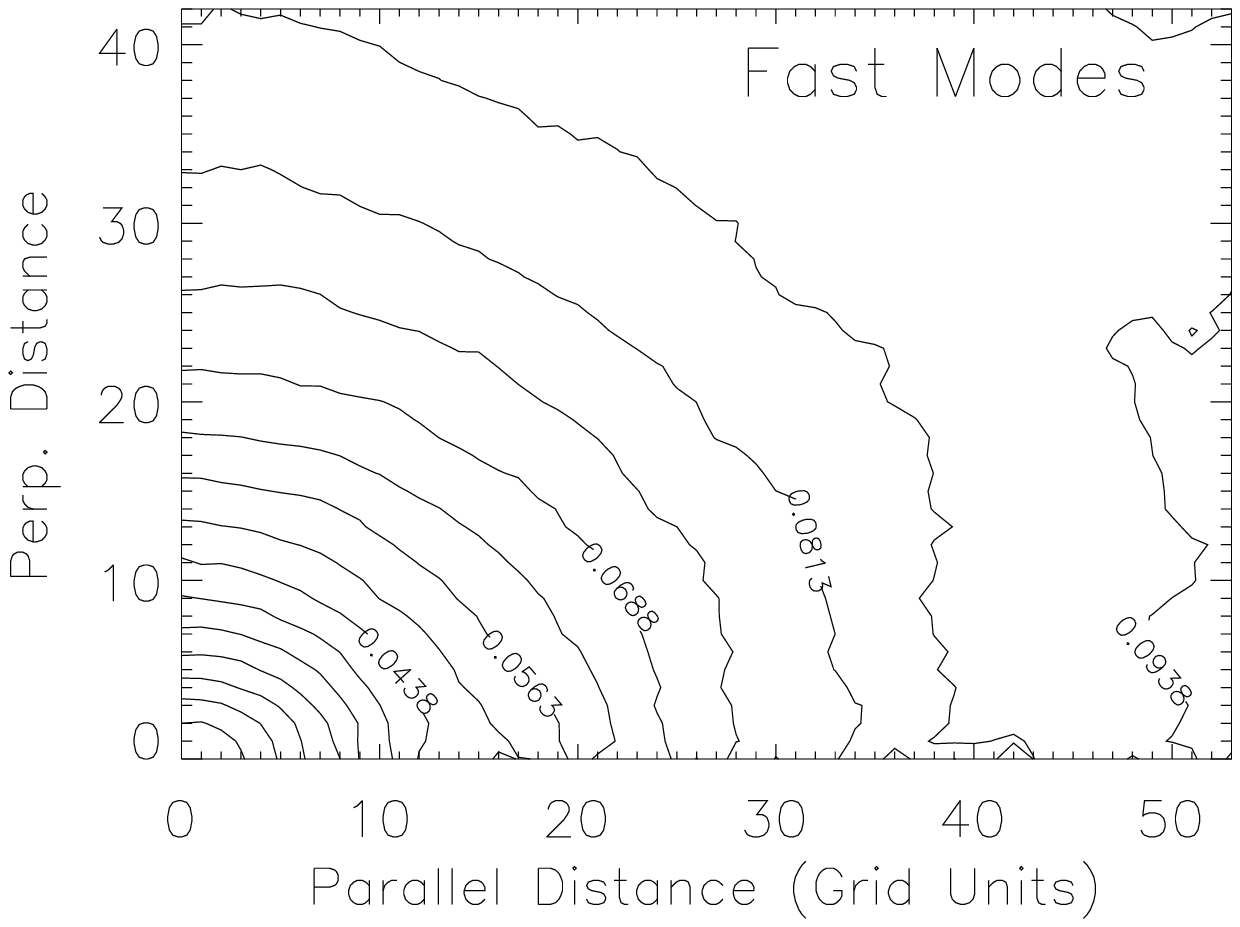}
\caption{
         (a) The power spectrum of fast waves is compatible with
             the IK spectrum.
         (b) The magnetic second-order structure function of 
             fast modes shows isotropy.{}From Cho \& Lazarian (2002a).
       }
\label{fig_fast}
\end{figure}

{}Figure \ref{fig_fast} shows fast modes are isotropic.
{}The resonance conditions for interacting fast waves are:
\begin{eqnarray}
\omega_1 + \omega_2 &=& \omega_3, \\ 
  {\bf k}_1 + {\bf k}_2 &=& {\bf k}_3.
\end{eqnarray}
Since $ \omega \propto k$ for the fast modes,
the resonance conditions can be met only when
all three ${\bf k}$ vectors are collinear.
This means that the direction of energy cascade is 
{\it radial} in Fourier space, and 
we expect an isotropic distribution of energy in Fourier
space.

Using the constancy of energy cascade and uncertainty principle, we
can derive an IK-like energy spectrum for fast waves.
The constancy of cascade rate reads
\begin{equation}
 \frac{ v_l^2 }{ t_{cas} } = \frac{ k^3v_k^2 }{ t_{cas} }=~\mbox{constant.}
\label{cas_rate_fast}
\end{equation}
On the other hand, $t_{cas}$ can be estimated as
\begin{equation}
t_{cas} \sim \frac{ v_k }{ \left( {\bf v}\cdot \nabla {\bf v} \right)_{\bf k} }
        \sim \frac{ v_k }{  \sum_{{\bf p}+{\bf q}={\bf k}}  kv_pv_q }.
\label{eq14_fast}
\end{equation}
If contributions are random,
the denominator can be written by the square root of the 
number of interactions ($\sqrt{\mathcal{N}}$)
times strength of individual interactions
($\sim kv_k^2$) \footnote{
To be exact, the strength of individual interactions is 
$\sim kv_k^2 \sin\theta$, where $\theta$ is the angle between ${\bf k}$ and
${\bf B}_0$. Thus marginal anisotropy is expected.
It will be investigated elsewhere.
}. 
Here we assume locality of interactions: $p\sim q\sim k$.
Due to the uncertainly principle,
the number of interactions becomes $\mathcal{N}\sim k(\Delta k)^2$,
where
$\Delta k$ is 
the typical transversal (i.e.~not radial) 
separation between two wave vectors ${\bf p}$ and ${\bf q}$ 
(with ${\bf p}+{\bf q}={\bf k}$).
Therefore, the denominator of equation (\ref{eq14_fast}) is 
$(k (\Delta k)^2)^{1/2} kv_k^2$.
We obtain an independent expression for $t_{cas}$ from 
the uncertainty principle ($t_{cas} \Delta \omega \sim 1$ 
with $\Delta \omega \sim \Delta k (\Delta k/k)$).
{}From this and equation (\ref{eq14_fast}), we get
$
t_{cas} \sim  t_{cas}^{1/2}/(k^{2} v_k),
$ 
which yields
\begin{equation}
t_{cas} \sim  1 / k^{4} v_k^2. \label{t_cascade}
\end{equation}
Combining equations (\ref{cas_rate_fast}) and (\ref{t_cascade}), we obtain
$ 
v_k^2 \sim  k^{-7/2},
$ 
or
$
 E^f(k) \sim k^2 v_k^2 \sim  k^{-3/2}$.
This is very similar to acoustic turbulence, turbulence caused by interacting
sound waves (Zakharov 1967; Zakharov \& Sagdeev 1970; L'vov, L'vov, \& Pomyalov 2000). 
Zakharov \& Sagdeev  (1970) found
$E(k)\propto k^{-3/2}$.
However, there is debate about
the exact scaling of acoustic turbulence.
Here we cautiously claim that our numerical results are compatible
with the Zakharov \& Sagdeev scaling:
\begin{equation}
\mbox{\it Spectrum of Fast Modes:~~~~~} E^f(k) \sim  k^{-3/2}.
\end{equation}

Magnetic field perturbations are mostly affected by
{}fast modes (Cho \& Lazarian 2002a) when $\beta$ is small:
\begin{equation} 
\mbox{\it Fast:~~~~~} \frac{(\delta B)_f}{ (\delta V)_A } 
  \sim  \frac{ (\delta V)_f }{ (\delta V)_A },
\end{equation}
if $(\delta V)_A \sim (\delta V)_s$.

The turbulent cascade of fast modes is expected to be slow
and in the absence of collisionless damping
they are expected to propagate in turbulent media
over distances considerably larger than Alfv\'en or
slow modes.
This effect is difficult to observe in
numerical simulations with $\Delta B \sim B_0$.
A modification of the spectrum in the presence of the collisionless
damping is presented in Yan \& Lazarian (2002).

\section{Implication for Cosmic Ray Propagation}

Many astrophysical problems require some knowledge of the scaling
properties of turbulence. Therefore we expect a wide range of applications
of the established scaling relations. Here we show how recent progress
in understanding MHD turbulence affects cosmic ray propagation.

The propagation of cosmic rays is mainly determined by their interactions
with electromagnetic fluctuations in interstellar medium. The resonant
interaction of cosmic ray particles with MHD turbulence has been repeatedly
suggested as the main mechanism for scattering and isotropizing cosmic
rays. In these analysis it is usually assumed that the turbulence
is \textit{isotropic} with a Kolmogorov spectrum (see Schlickeiser
\& Miller 1998). How should these calculations be modified?

Consider resonance interaction first. Particles moving with velocity $ v $ get into resonance
with MHD perturbations propagating along the magnetic field if the
resonant condition is fulfilled, namely, $\omega =k_{\parallel }v\mu +n\Omega $
($n=\pm 1,2...$), where $\omega $ is the wave frequency, $\Omega =\Omega _{0}/\gamma $
is the gyrofrequency of relativistic particle, $\mu =\cos \alpha $,
where $\alpha $ is the pitch angle of particles. In other words,
resonant interaction between a particle and the transverse electric
field of a wave occurs when the Doppler shifted frequency of the wave
in the particle's guiding center rest frame $\omega _{gc}=\omega -k_{\parallel }v\mu $
is a multiple of the particle gyrofrequency.

For cosmic rays, $k_{\parallel }v\mu\gg \omega $, so the
slow variation of the magnetic field with time can be neglected. Thus
the resonant condition is simply $k_{\parallel }v\mu +n\Omega =0$.
From this resonance condition, we know that the most important interaction
occurs at $k_{\parallel }=k_{res}=\Omega /v_{\parallel }$.

It is intuitively clear that resonant interaction of particles in
isotropic and anisotropic turbulence should be different. Chandran
(2001) obtained strong suppression of scattering by Alfvenic turbulence
when he treated turbulence anisotropies in the spirit of Goldreich-Sridhar
model of incompressible turbulence. His treatment was improved in
Yan \& Lazarian (2002, henceforth YL02) who used a more rigorous description
of magnetic field statistics. Moreover, they took into account CR scattering by compressible MHD modes and found that fast modes absolutely dominate cosmic ray scattering. In our description we shall follow YL02 treatment of
the problem.

We employ quasi-linear theory (QLT) to obtain our estimates. QLT has
been proved to be a useful tool in spite of its intrinsic limitations
(Chandran 2000; Schlickeiser \& Miller 1998; Miller 1997). For moderate
energy cosmic rays, the corresponding resonant scales are much smaller
than the injection scale. Therefore the fluctuation on the resonant
scale $\delta B\ll B_{0}$ even if they are comparable at the injection
scale. QLT disregards diffusion of cosmic rays that follow wandering
magnetic field lines (Jokipii 1966) and this diffusion should be accounted
separately. Obtained by applying the QLT to the collisionless Boltzmann-Vlasov
equation, the Fokker-Planck equation is generally used to describe
the evolvement of the gyrophase-average distribution function $f$,

\[
\frac{\partial f}{\partial t}=\frac{\partial }{\partial \mu }\left(D_{\mu \mu }\frac{\partial f}{\partial \mu }+D_{\mu p}\frac{\partial f}{\partial p}\right)+\frac{1}{p^{2}}\frac{\partial }{\partial p}\left[p^{2}\left(D_{\mu p}\frac{\partial f}{\partial \mu }+D_{pp}\frac{\partial f}{\partial p}\right)\right],\]
 where $p$ is particle momentum. The Fokker-Planck coefficients $D_{\mu \mu },D_{\mu p},D_{pp}$
are the fundamental physical parameter for measuring the stochastic
interactions, which are determined by the electromagnetic fluctuations
(Schlickeiser \& Achatz 1993). From Ohm's Law $\mathbf{E}(\mathbf{k})=-(1/c)\mathbf{v}(\mathbf{k})\times \mathbf{B}_{0},$
we can get the electromagnetic fluctuations from correlation tensors
of magnetic and velocity fluctuations $C_{ij},$ $K_{ij}$. Here,
\begin{eqnarray}
<B_{i}(\mathbf{k})B_{j}^{*}(\mathbf{k'})>/B_{0}^{2}=\delta (\mathbf{k}-\mathbf{k'})M_{ij}(\mathbf{k}), &  & \nonumber \\
<v_{i}(\mathbf{k})B_{j}^{*}(\mathbf{k'})>/V_{A}B_{0}=\delta (\mathbf{k}-\mathbf{k'})C_{ij}(\mathbf{k}), &  & \nonumber \\
<v_{i}(\mathbf{k})v_{j}^{*}(\mathbf{k'})>/V_{A}^{2}=\delta (\mathbf{k}-\mathbf{k'})K_{ij}(\mathbf{k}). &  & 
\end{eqnarray}
 
For Alfven modes, Cho, Lazarian and Vishniac (2002) obtained \begin{equation}
K_{ij}(\mathbf{k})=C_{a}I_{ij}k_{\perp }^{-10/3}exp(-L^{1/3}k_{\parallel }/k_{\perp }^{2/3}),\label{anisotropic}\end{equation}
 where $I_{ij}=\{\delta _{ij}-k_{i}k_{j}/k_{\perp }^{2}\}$ is a 2D
matrix in x-y plane, $k_{\parallel }$ is the wave vector along the
local mean magnetic field, $k_{\perp }$ is the wave vector perpendicular
to the magnetic field and the normalization constant $C_{a}=L^{-1/3}/6\pi $.
We assume that for the Alfven modes $M_{ij}=K_{ij},$ $C_{ij}=\sigma M_{ij}$
where the fractional helicity $-1<\sigma <1$ is independent of $\mathbf{k}$
(Chandran 2000). According to Cho \& Lazarian (2002a), fast modes
are isotropic and have one dimensional spectrum $E(k)\propto k^{-3/2}$.
In low $\beta $ medium, the velocity fluctuations are always perpendicular
to $\mathbf{B}_{0}$ for all $\mathbf{k}$, while the magnetic fluctuations
are perpendicular to $\mathbf{k}$. Thus $K_{ij},$ $M_{ij}$ of fast
modes are not equal, \begin{equation}
\left[\begin{array}{c}
 M_{ij}({\mathbf{k}})\\
 C_{ij}({\mathbf{k}})\\
 K_{ij}({\mathbf{k}})\end{array}
\right]={\frac{L^{-1/2}}{8\pi }}J_{ij}k^{-7/2}\left[\begin{array}{c}
 \cos ^{2}\theta \\
 \sigma \cos ^{2}\theta \\
 1\end{array}
\right],\label{fast_tensor_lowb}\end{equation}
 where $J_{ij}=k_{i}k_{j}/k_{\perp }^{2}$ is also a 2D tensor in
$x-y$ plane %
\footnote{Apparently $M_{ij},$ $C_{ij}$ are 3D matrixes. However, the third
dimension is not needed for our calculations. $M_{ij}$ is different
from that in Schlickeiser \& Miller (1998). The fact that the fluctuations
$\delta \mathbf{B}$ in fast modes are in the $\mathbf{k}$-$\mathbf{B}$
plane place another constrain on the tensor so that the term $\delta _{ij}$
doesn't exist.%
}. In high $\beta $ medium, the velocity fluctuations are radial,
i.e., along the direction of ${\textbf {k}}$. Fast modes in this
regime are essentially sound waves compressing magnetic field (GS95;
Lithwick \& Goldreich 2001, Cho \& Lazarian, in preparation). The
compression of magnetic field depends on plasma $\beta $. The corresponding
x-y components of the tensors are \begin{equation}
\left[\begin{array}{c}
 M_{ij}({\mathbf{k}})\\
 C_{ij}({\mathbf{k}})\\
 K_{ij}({\mathbf{k}})\end{array}
\right]={\frac{L^{-1/2}}{8\pi }}\sin ^{2}\theta J_{ij}k^{-7/2}\left[\begin{array}{c}
 \cos ^{2}\theta /\beta \\
 \sigma \cos \theta /\beta ^{1/2}\\
 1\end{array}
\right].\label{fast_tensor_highb}\end{equation}
 Adopting the approach in Schlickeiser \& Achatz (1993), we can obtain
the Fokker-Planck coefficients in the lowest order approximation of
$V_{A}/c$, \begin{eqnarray}
\left[\begin{array}{c}
 D_{\mu \mu }\\
 D_{\mu p}\\
 D_{pp}\end{array}
\right]={\frac{\Omega ^{2}(1-\mu ^{2})}{2B_{0}^{2}}}\left[\begin{array}{c}
 1\\
 mc\\
 m^{2}c^{2}\end{array}
\right]{\mathcal{R}}e\sum _{n=-\infty }^{n=\infty }\int _{k_{min}}^{k_{max}}dk^{3} &  & \nonumber \\
\int _{0}^{\infty }dte^{-i(k_{\parallel }v_{\parallel }-\omega +n\Omega )t}\left\{ J_{n+1}^{2}({\frac{k_{\perp }v_{\perp }}{\Omega }})\left[\begin{array}{c}
 P_{{\mathcal{RR}}}({\mathbf{k}})\\
 T_{{\mathcal{RR}}}({\mathbf{k}})\\
 R_{{\mathcal{RR}}}({\mathbf{k}})\end{array}
\right]\right. &  & \nonumber \\
+J_{n-1}^{2}({\frac{k_{\perp }v_{\perp }}{\Omega }})\left[\begin{array}{c}
 P_{{\mathcal{LL}}}({\mathbf{k}})\\
 -T_{{\mathcal{LL}}}({\mathbf{k}})\\
 R_{{\mathcal{LL}}}({\mathbf{k}})\end{array}
\right]+J_{n+1}({\frac{k_{\perp }v_{\perp }}{\Omega }})J_{n-1}({\frac{k_{\perp }v_{\perp }}{\Omega }}) &  & \nonumber \\
\left.\left[e^{i2\phi }\left[\begin{array}{c}
 -P_{{\mathcal{RL}}}({\mathbf{k}})\\
 T_{{\mathcal{RL}}}({\mathbf{k}})\\
 R_{{\mathcal{RL}}}({\mathbf{k}})\end{array}
\right]+e^{-i2\phi }\left[\begin{array}{c}
 -P_{{\mathcal{LR}}}({\mathbf{k}})\\
 -T_{{\mathcal{LR}}}({\mathbf{k}})\\
 R_{{\mathcal{LR}}}({\mathbf{k}})\end{array}
\right]\right]\right\}  &  & \label{genmu}
\end{eqnarray}
 where $k_{min}=L^{-1}$, $k_{max}=\Omega _{0}/v_{th}$ corresponds
to the dissipation scale, $m=\gamma m_{H}$ is the relativistic mass
of the proton, $v_{\perp }$ is the particle's velocity component
perpendicular to $\mathbf{B}_{0}$, $\phi =\arctan (k_{y}/k_{x}),$
${\mathcal{L}},{\mathcal{R}}=(x\pm iy)/\sqrt{2}$ represent left and
right hand polarization%
\footnote{For $D_{\mu p}$, the expression is only true for Alfv\textbackslash{}'en
modes. There are additional compressional terms for compressable modes.%
}.

\subsection{Scattering by Alfvenic turbulence}

Noticing that the integrand for small $k_{\perp }$ is substantially
suppressed by the exponent in the anisotropic tensor (see Eq. (\ref{anisotropic}))
so that the large scale contribution is not important, we can simply
use the asymptotic form of Bessel function for large argument. Then
if the pitch angle $\alpha $ is not close to 0, we can derive the
analytical result for anisotropic turbulence (YL02), \begin{equation}
\left[\begin{array}{c}
 D_{\mu \mu }\\
 D_{\mu p}\\
 D_{pp}\end{array}
\right]=\frac{v^{2.5}\cos \alpha ^{5.5}}{2\Omega ^{1.5}L^{2.5}\sin \alpha }\Gamma [6.5,k_{max}^{-2/3}k_{res}L^{1/3}]\left[\begin{array}{c}
 1\\
 \sigma mV_{A}\\
 m^{2}V_{A}^{2}\end{array}
\right],\label{ana}\end{equation}
 where $L$ is the injection sale, $k_{max}=\Omega _{0}/v_{th}$ corresponds
to the dissipation scale, $\Gamma [a,z]$ represents the incomplete
gamma function. 

\begin{figure}
\resizebox*{0.49\columnwidth}{!}{\includegraphics{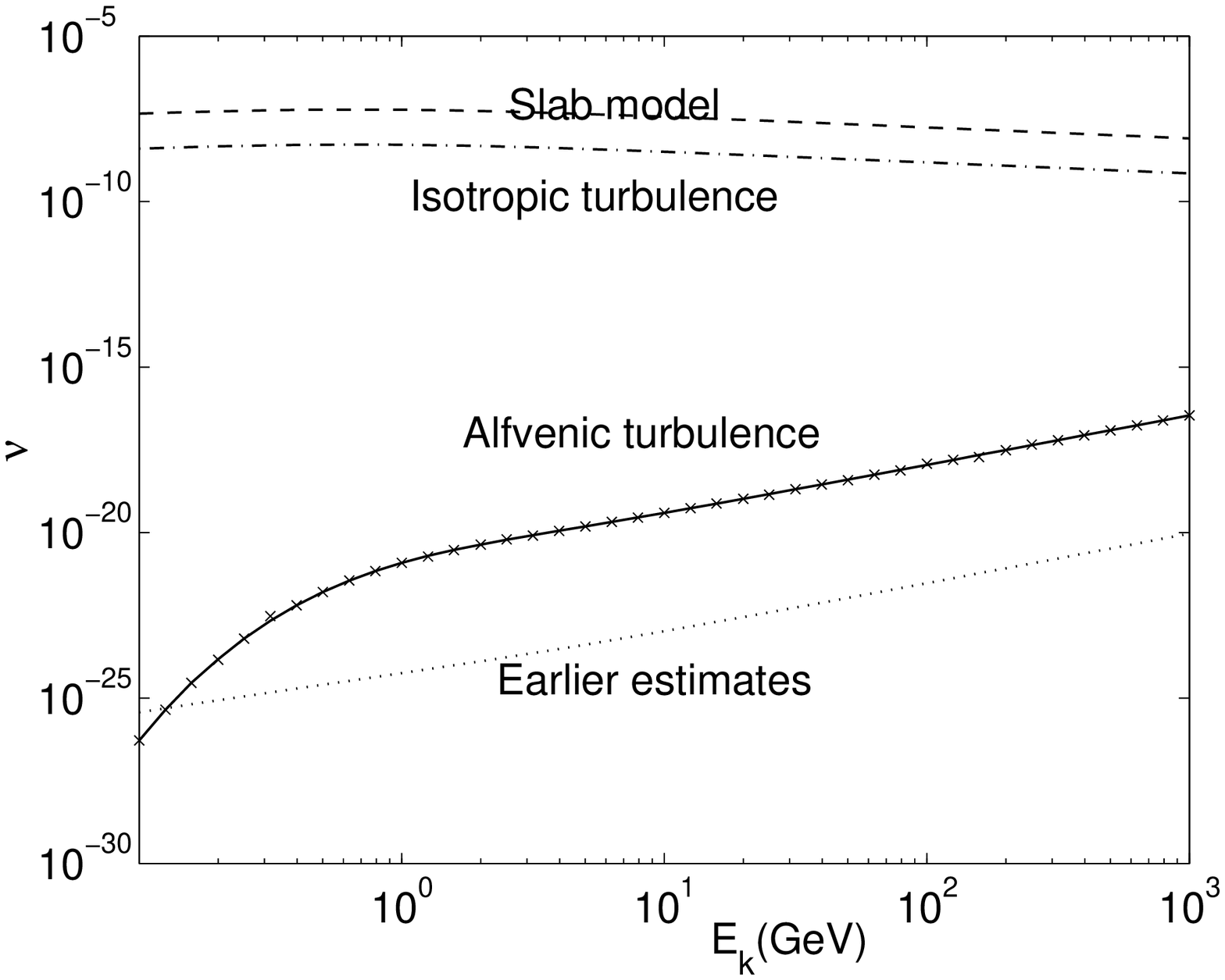}} \hfil
\resizebox*{0.49\columnwidth}{!}{\includegraphics{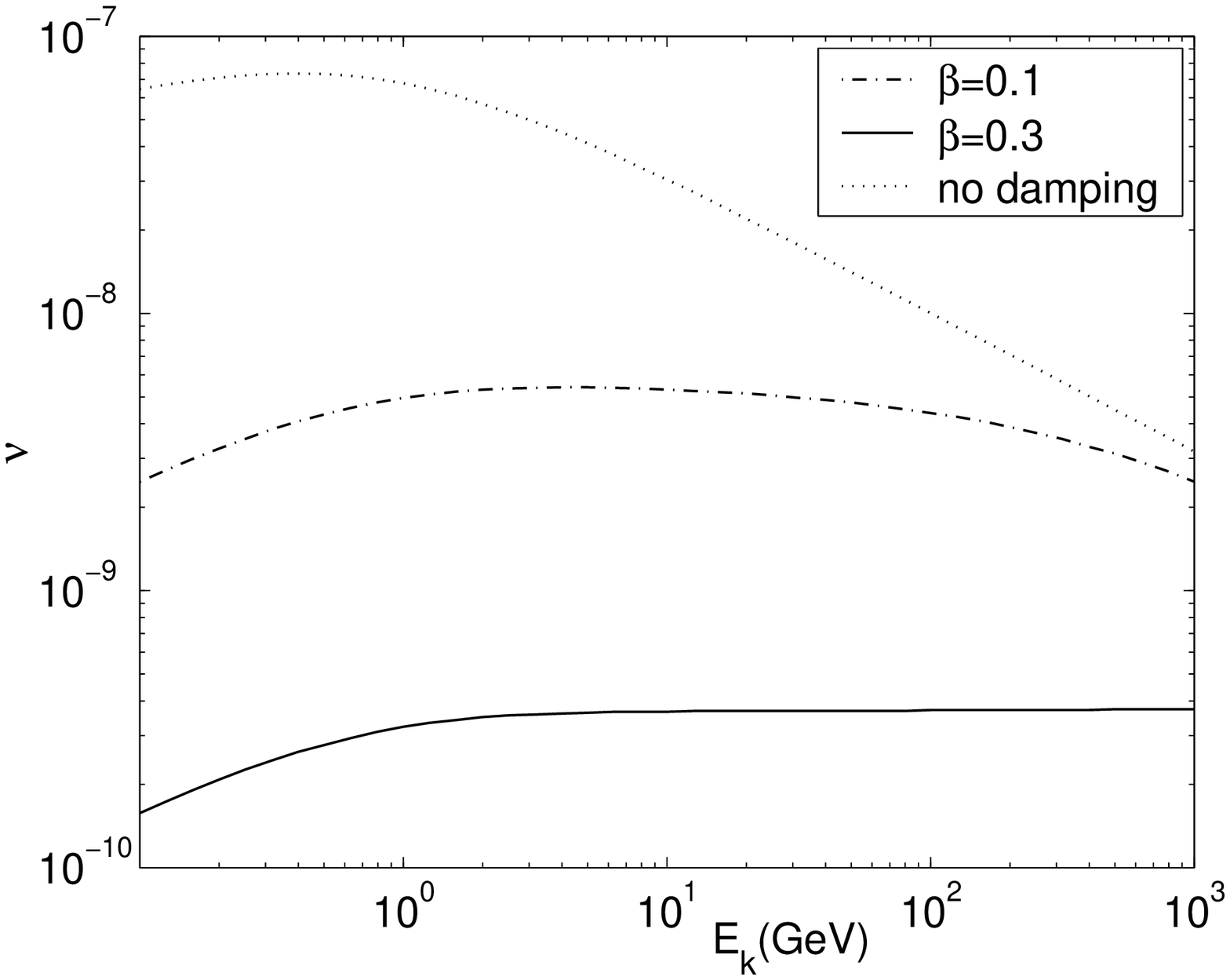}}%
\caption{The scattering frequency \protect$\nu $ vs. the kinetic energy
\protect$E_{k}$ of cosmic rays (a) by Alfv\'{e}nic turbulence,
(b) by fast modes. In (a), the dash-dot line refers to the scattering
frequency for isotropic turbulence. The '\protect\protect\protect\protect\protect$\times $'
represents our numerical result for anisotropic turbulence, the solid
line is our analytical result from Eq.~(\ref{ana}). Also plotted
(dashed line) is the previous result for anisotropic turbulence in
Chandran (2001). In (b), the dashed line represents the scattering
by fast modes not subjected to damping, the solid and dashdot line are
the results taking into account collisionless damping. (From Yan \&
Lazarian 2002)}

\label{fig:incom}
\end{figure}

The scattering frequency $\nu =2D_{\mu \mu }/(1-\mu ^{2})$ is plotted
for different models in Fig.(\ref{fig:incom}a). It is clear that
anisotropy suppresses scattering. Although our results are larger
than those obtained in Chandran (2001) using an \textit{ad hoc} tensor
with a step function%
\footnote{We counted only the resonant term in Chandran (2001). The nonresonant
term is spurious as noted in Chandran (2001)%
}, they are still much smaller than the estimates for isotropic model.
Unless we consider very high energy CRs ($\geq 10^{8}GeV$) with corresponding
Larmor radius comparable to the injection scale, we can neglect scattering
by the Alfv\'{e}nic turbulence. What is the alternative way to scatter
cosmic rays?

\subsection{Scattering by fast modes}

For compressible modes we discuss two types of resonant interaction:
gyroresonance and transit-time damping; the latter requires longitudinal
motions. The contribution from slow modes is not larger than that
by Alfv\'{e}n modes since the slow modes have the similar anisotropies
and scalings. More promising are fast modes, which are isotropic (Cho
\& Lazarian 2002a). However, fast modes are subject to collisionless
damping if the wavelength is smaller than the proton mean free path
or by viscous damping if the wavelength is larger than the mean free
path. According to CL02, fast modes cascade over time scales $\tau _{fk}=\tau _{k}\times V_{A}/v_{k}=(k\times k_{min})^{-1/2}\times V_{A}/V^{2},$
where $\tau _{k}=kv_{k}$ is the eddy turn-over time, $V$ is the
turbulence velocity at the injection scale.

Consider gyroresonance scattering in the presence of collisionless
damping. The cutoff of fast modes corresponds to the scale where $\tau _{fk}\gamma _{d}\simeq 1$
and this defines the cutoff scale $k_{c}^{-1}$. Using the tensors
given in Eq.~(\ref{fast_tensor_lowb}) we obtain the corresponding
Fokker-Planck coefficients for the CRs interacting with fast modes
by integrating Eq.(\ref{genmu}) from $k_{min}$ to $k_{c}$ (see
Fig.(\ref{fig:incom}b)). When $k_{c}^{-1}$ is less than $r_{L}$,
the results of integration for damped and undamped turbulence coincides.
Since the damping increases with $\beta $, the scattering frequency
decreases with $\beta $.

Adopting the tensors given in Eq.~(\ref{fast_tensor_highb}), it is possible to calculate
the scattering frequency of CRs in high $\beta $ medium. For instance,
for density $n=0.5$cm$^{-3},$ temperature $T=8000$K, magnetic field
$B_{0}=1\mu $G, the mean free path is smaller than the resonant wavelength
for the particles with energy larger than $0.1GeV$, therefore collisional
damping rather than Landau damping should be taken into account. Nevertheless,
our results show that the fast modes still dominate the CRs' scattering
in spite of the viscous damping.

Apart from the gyroresonance, fast modes potentially can scatter CRs
by transit-time damping (TTD) (Schlickeiser \& Miller 1998). TTD happens due to
the resonant interaction with parallel magnetic mirror force $-(mv_{\perp }^{2}/2B)\nabla _{\parallel }\mathbf{B}$.
For small amplitude waves, particles should be in phase with the wave
so as to have a secular interaction with wave. This gives the Cherenkov
resonant condition $\omega -k_{\parallel }v_{\parallel }\sim 0$,
corresponding to the $n=0$ term in Eq.(\ref{genmu}). From the condition,
we see that the contribution is mostly from nearly perpendicular propagating
waves ($\cos \theta \sim 0$). According to Eq.~(\ref{fast_tensor_lowb}),we see that the
corresponding correlation tensor for the magnetic fluctuations $M_{ij}$
are very small, so the contribution from TTD to scattering is not
important.

Self-confinement due to the streaming instability has been discussed
by different authors(see Cesarsky 1980, Longair 1997) as an effective alternative
to scatter CRs and essential for CR acceleration by shocks. However,
we will discuss in our next paper that in the presence of the turbulence
the streaming instability will be partially suppressed owing to the
nonlinear interaction with the background turbulence.

Thus the gyroresonance with the fast modes is the principle mechanism
for scattering cosmic rays. This demands a substantial revision of
cosmic ray acceleration/propagation theories, and many related problems
may need to be revisited. For instance, our results may be relevant
to the problems of the Boron to Carbon abundances ratio. We shall
discuss the implications of the new emerging picture elsewhere.

\section{Summary}

Recently there have been significant advances in the field of compressible
MHD turbulence and its implications to cosmic ray transport:

\begin{enumerate}
\item Simulations of compressible MHD turbulence show that there is a weak
coupling between Alfv\'{e}n waves and compressible MHD waves and
that the Alfv\'{e}n modes follow the Goldreich-Sridhar scaling. Fast
modes, however, decouple and exhibit isotropy. 
\item Scattering of cosmic rays by Alfvenic modes is suppressed and therefore
the scattering by fast modes is the dominant process provided that
turbulent energy is injected at large scales. 
\item The scattering frequency by fast modes depends on collisionless damping
or viscous damping and therefore on plasma $\beta $. 
\end{enumerate}
\textbf{Acknowledgments:} We acknowledge the support of the NSF through
grant AST-0125544. This work was partially supported by National Computational
Science Alliance under AST000010N and utilized the NCSA SGI/CRAY Origin2000.

\end{document}